\documentclass[%
 reprint,
 amsmath,amssymb,
 aps,
 prb,
 floatfix,
]{revtex4-1}

\usepackage{dcolumn}
\usepackage{bm}
\usepackage{grffile}
\usepackage{siunitx}

\usepackage{footmisc}
\usepackage{setspace}
\usepackage{braket}
\usepackage{graphicx}
\usepackage{enumitem}

\renewcommand{\vec}[1]{\boldsymbol{#1}}
\newcommand{\mat}[1]{\boldsymbol{#1}}
\newcommand{\op}[1]{\hat{#1}}
\newcommand{\doubleunderline}[1]{\underline{\underline{#1}}}

\begin{document}

\preprint{APS/123-QED}

\title{Clean and Convenient Tessellations for Number Counting Jastrow Factors}


\author{Brett Van Der Goetz$^1$}
\author{Leon Otis$^1$}
\author{Eric Neuscamman$^{1,2,}$}
 \email{eneuscamman@berkeley.edu}

\affiliation{
${}^1$Department of Chemistry, University of California, Berkeley, CA, 94720, USA \\
${}^2$Chemical Sciences Division, Lawrence Berkeley National Laboratory, Berkeley, CA, 94720, USA
}


\date{\today}

\begin{abstract}
We introduce a basis of counting functions that, by cleanly tessellating
three dimensional space, allows real space number counting Jastrow factors
to be straightforwardly applied to general molecular situations.
By exerting direct control over electron populations in local
regions of space and encoding pairwise correlations between these
populations, these Jastrow factors allow even very simple reference
wave functions to adopt nodal surfaces well suited to many strongly
correlated settings.
Being trivially compatible with traditional Jastrow factors and
diffusion Monte Carlo and having the same cubic per-sample cost
scaling as a single determinant trial function,
these Jastrow factors thus offer a powerful new route to
the simultaneous capture of weak and strong electron correlation
effects in a wide variety of molecular and materials settings.
In multiple strongly correlated molecular examples, we show that
even when paired with the simplest
possible single determinant reference, these Jastrow factors
allow quantum Monte Carlo to
out-perform coupled cluster theory and approach the accuracy of
traditional multi-reference methods.
\end{abstract}

\maketitle

\section{Introduction}
\label{sec:intro}

Characterizing novel strongly-correlated systems requires an accurate description of the electronic wavefunction.
Traditionally, this is achieved by adding correlations to a minimally-correlated mean-field reference\cite{helgaker,szabo_ostlund}.
It is often useful to divide these correlations into two distinct classes -- strong (static) and weak (dynamic)\cite{strong_weak_correlation} -- based on their relative magnitudes and impact on the accuracy of the reference wavefunction.
This dichotomy is not well-defined, but it is useful in limiting cases, as each class describes effects that arise in different physical limits. 
As different types of these correlation effects take on distinct mathematical forms, they are often treated using distinct theoretical methods.
A quantitative method must consider both to a high accuracy, a task that sharply grows in difficulty with the complexity and size of the system.

These corrections introduce statistical correlations between individual electrons within a many-body reference wavefunction.
Static or strong correlations are those that require large, qualitative corrections to a mean-field state to effectively describe, and manifest in chemical systems as stretched or broken bonds.
By contrast, weak or dynamic correlations are those that give rise to more subtle effects like particle coalescence cusps\cite{kato_cusps,needs_cusp,qmc_review} and van der Waals interactions\cite{parsegian_vdw}, and are often described by perturbative or explicitly correlated\cite{helgaker,valeev_f12} methods.
Without additional approximations, combining methods individually well-suited for different classes of correlations can lead to prohibitive computational costs that are much greater than the cost of the components.
As a result, theoretical methods in quantum chemistry still struggle to capture important contributions from both types of correlations simultaneously in complex systems while remaining computationally tractable\cite{pulay_active_space, chromium_dimer,oec_dmrg, lee2018open}.

In Fock space, static correlation is often treated using active-space methods\cite{hartree_casscf,roos_casscf,rasscf}, and involves optimizing the energy of a wavefunction restricted to a subspace of configurations generated from orbitals within a limited energy window.
In this subspace, the number of configurations scales combinatorially with the number of electrons and orbitals, and in cases where highly degenerate patterns of orbitals emerge, even the size of this active space becomes impractical.
Although great progress has been made in addressing this challenge by  DMRG\cite{white_dmrg2,white_dmrg,oec_dmrg,chan_dmrg,chan_dmrg2,kawakami_dmrg}, selective CI\cite{huron_sci,evangelisti_cipsi,kawakami_dmrg,sharma_hci,sharma_hci2,giner_scidmc,giner_scidmc2}, and FCIQMC\cite{alavi_fciqmc,alavi_fciqmc_nature,foulkes_fciqmc}, it remains difficult to go beyond 40 active orbitals due to the exponential asymptotic scaling that these methods all possess.

Dynamic correlation is typically captured using perturbation theory\cite{szabo_ostlund}.
These corrections are vital for accuracy, as they capture details such as wavefunction cusps\cite{kato_cusps} and van der Waals correlations\cite{parsegian_vdw}.
However, the cost of including these corrections scales both with the system size and the size of the already-complex active-space wavefunction, and so in practice, adding dynamic correlation effects on top of aggressive static correlation treatments remains quite challenging.
For example, in approaches like CASPT2\cite{andersson_caspt2}, the cost of perturbatively adding dynamic correlation to a CAS reference scales as the sixth power of the size of the active space.
As a result, it remains interesting to look for approaches that incorporate static and dynamic correlation at low-order polynomial cost, even if they may be more approximate than other post-CAS methods.

By contrast, real-space wavefunctions offer a number of powerful advantages when attempting to capture both types of correlation simultaneously.
For instance, the Slater-Jastrow wavefunction\cite{qmc_review} is natively attuned to wavefunction cusps and can exactly express them using a small set of variational parameters.
The Jastrow factor, a symmetric many-body multiplicative factor, is responsible for this compact description, as electron-electron cusps are naturally described using the relative particle coordinates available to its explicit position-space representation.
This is in stark contrast to Fock space, where cusps may not have an exact representation and where an equivalent factor would require a lengthy enumeration of corrections\cite{helgaker,valeev_f12}. 
This choice between real-space and Fock space wavefunctions matters, as concise descriptions of particle correlations may appear only within a particular representation.
Those acquainted with real-space Monte Carlo\cite{qmc_review} techniques might argue that projector-based methods such as Diffusion Monte Carlo (DMC) obviate the explicit formulation of every detail of correlation, a powerful advantage not enjoyed by many advanced Fock space methods. 
However, these methods rely on the accuracy of the wavefunction's nodal surface, which, though determined exclusively by the wavefunction's antisymmetric components, is nonetheless coupled to the symmetric Jastrow factor.

There have been efforts to expand the scope of Jastrow factors beyond particle cusp by writing three-body and four-body correlation terms 
using higher-order functions of interparticle coordinates\cite{lester_qmc,nightingale_jastrow,mood_jastrow,needs_jastrow}
or a standard atomic orbital basis\cite{sorella_jastrow4,sorella_jastrow5,sorella_jastrow,beaudet_jastrows,guidoni_jastrow,sorella_jastrow2,guidoni_jastrow2,sorella_jastrow3}.
Although the hydrogenic functions used in this expansion have proven successful building blocks for these sophisticated Jastrow factors, they have some formal shortcomings that makes the exploration of alternatives worthwhile, especially in the context of strong correlation.
For example, enacting local projections on electron number, which can be useful when treating static correlation, requires a very large basis of hydrogenic functions due to the fact that the basis elements have non-zero curvature near their centers.
As number projections can help break chemical bonds\cite{neuscamman_cjagp}, a Jastrow basis that cleanly accommodates them hold promise as a way to compactly encode some static correlation in a way that complements existing Jastrow factors.

In contrast to most standard methods, where a wavefunction is constructed through a hierarchy of additive corrections, projection factors remove high-energy components already present in a reference state.
Number projections are particularly suited to stenciling out these high-energy components from compact functional forms which contain an overabundance of spurious ionic configurations, such as the antisymmeterized geminal power (AGP) wavefunction\cite{linderberg_agp}. 
Application of a Gutzwiller-style \cite{gutzwiller_gf} number projection factor has been shown to successfully restore size consistency to the AGP\cite{neuscamman_jagp}, a feat which would otherwise require an exponential number of individual additive corrections.

We have recently shown\cite{vdg_ncjf} that Jastrow factors with a basis of sigmoidal ``counting functions'' -- real space functions which seek to mimic Fock space number operators -- successfully perform this number projection when applied to single-determinant reference wavefunctions.
Not only are these factors trivially compatible with existing Jastrow formulations that help treat dynamic correlation, but their ability to projectively encode static correlations may also help mitigate the size of multi-reference determinantal expansions needed to achieve accurate predictions.
However, the real-space counting functions we used previously were constructed in an \emph{ad hoc} way, and were not straightforwardly extensible to general chemical systems.

In this paper, we present a Number counting Jastrow Factor that is capable of number projection using a basis of automatically-constructed counting functions.
These counting functions partition space into disjoint regions and tally the total electron count within each, and pairs of the resulting populations are correlated in an exponential Jastrow factor.
The functional form chosen for these counting functions provides them both with the flexibility to adopt complex, quadratic shapes and the simple formal properties that permit the creation of near-arbitrary spatial divisions.
We provide two straightforward and automatic generation schemes for these counting functions which can be used to partition space into either spherical sections or polyhedral cells.
These Jastrow factors were tested by applying them to single-determinant reference wavefunctions in several strongly correlated systems and exhibit surprising accuracy despite the simplicity of the reference state.
In the future, these Jastrows may be combined with more sophisticated fermionic components such as multi-Slater expansions or geminal power wavefunctions, but in this study we restrict ourselves to pairing them with a single determinant in order to explore just how far they can take us towards polynomial-cost static correlation on their own.

\section{Theory}

\subsection{Review}
\label{sec-review}
\begin{table}
\centering
\caption{Key for mathematical notation.}
\begin{tabular}{c|c}
\text{Symbol} & \text{Description} \\
\hline
$n_e$ & Number of electrons \\
$n_C$ & Number of counting functions \\
$i,j,\dots$ & Particle index \\
$I,J,\dots$ & Counting region index \\ 
$\alpha, \beta, \dots$ & Atomic index \\
$a,b,\dots$ & Molecular orbital index \\
$\vec{r}_i$ & Single particle position, $\in \mathbb{R}^3$ \\
$\{\vec{r}_i\}$ & Particle configuration, $\in \mathbb{R}^{3n_e}$ \\
$N_I$ & Region population \\
$\mat{F}_{IJ}, \vec{G}_I$ & Population correlation coefficients \\
$C_I$ & Counting function \\
$\mathcal{R}_I$ & Counting region, $\subset\mathbb{R}^3$ \\
$\partial \mathcal{R}_I$ & Boundary of $\mathcal{R}_I$ \\
$\mathcal{V}_I$ & Voronoi region, $\subset \mathbb{R}^3$ \\
$C_{I/J}$ & Pairwise counting function \\
$\mathcal{R}_{I/J}$ & Pairwise counting region \\
$\mathcal{V}_{I/J}$ & Pairwise Voronoi region \\
$g_I$ & Anchor gaussian \\
$\vec{n}$ & Normal vector \\
$\vec{\mu}$ & Mean vector \\
$\rho_I(N)$ & Population density of $\mathcal{R}_I$ \\
\end{tabular}
\end{table}

Number counting Jastrow Factors, or counting Jastrows, are symmetric many-body factors defined in terms of region populations $N_I$ and linear coefficients $\mat{F}_{IJ}, \vec{G}_{I}$:
\begin{equation}\label{eq-ncjf_def} e^{J_{C}} = \exp\left( \sum\limits_{IJ} \mat{F}_{IJ} N_{I} N_{J} + \sum\limits_{I} \vec{G}_I N_I \right) \end{equation}
Each region population $N_I$ estimates the total electron population in a spatial region $\mathcal{R}_I$ for each particle configuration.
Counting Jastrow factors are able to perform particle-number projections between these regions which can be directly seen by transforming the region populations $\{N_I\}$ to the basis that diagonalizes $\mat{F}$:
\begin{equation} e^{J_C} = \exp\left( \sum\limits_{I} D_{II} \left( P_I - \tilde{N}_I\right)^2  \right), \quad \mat{F} = \mat{U}\mat{D}\mat{U}^{T} \end{equation}
When $D_{II}$ is negative, particle configurations are suppressed when the transformed region populations $\tilde{N}_I$ deviate from the prescribed population $P_I$, similar to performing a Hilbert space number projection\cite{neuscamman_jagp,neuscamman_cjagp}.
Region populations are calculated using a set of real-space single-particle counting functions $\{C_I\}$ which are designed to behave like indicator functions over the counting regions $\{\mathcal{R}_I\}$:
\begin{equation}\label{eq-cf_def} C_I(\vec{r}) \approx \left\{ \begin{array}{cc} 1 & \vec{r} \in \mathcal{R}_I \\ 0 & \vec{r} \not \in \mathcal{R}_I  \end{array} \right. \end{equation}
As these functions are contained within the Jastrow factor, we must continuously approximate this discrete switch at the counting region boundary to prevent singularities from appearing in the wavefunction gradient.
Each population is determined by summing the counting function values evaluated at each particle coordinate $\vec{r}_i$:
\begin{equation} N_I = \sum\limits_i C_I(\vec{r}_i) \end{equation}
The primary focus of this section is to detail a simple and flexible functional form for these counting functions with well-characterized counting regions that can be automatically constructed in a fully \emph{ab initio} way.

We had previously\cite{vdg_ncjf} used simple counting functions of the form:
\begin{equation} C(\vec{r}) = \frac{1}{1 + \exp\left( \alpha f(\vec{r}) \right)}\end{equation}
which is a sigmoidal function with an inflection point at the zeros of $f$, at which it attains a maximum slope proportional to $\alpha$. 
In the infinite slope limit, we can directly relate this to equation \ref{eq-cf_def}:
\begin{equation} \lim\limits_{\alpha\rightarrow \infty} \frac{1}{1+\exp\left(\alpha f(\vec{r}) \right)} = \left\{\begin{array}{cc} 1 & f(\vec{r}) < 0 \\ 1/2 & f(\vec{r}) = 0 \\ 0 & f(\vec{r}) > 0 \end{array} \right. \end{equation}
and we find that the counting regions are totally determined by the sign-structure of the function $f$:
\begin{equation} \mathcal{R} = \{ \vec{r}: f(\vec{r}) < 0 \}\end{equation}
The counting region boundary -- which we will refer to as the switching surface, denoted $\partial\mathcal{R}$ -- is the set of zeros of $f$ or, equivalently, when the counting function attains the intermediate value of one-half:
\begin{equation} \partial \mathcal{R} = \{ \vec{r}: f(\vec{r}) = 0 \} = \{\vec{r}: C(\vec{r}) = 1/2 \}\end{equation}
By implicitly defining the switching surface through the zeros of $f$, we avoid committing it to any particular shape or topology.
This approach is general and flexible, and mirrors techniques used in level set methods used to describe and propagate interfaces in fluid dynamics, computational geometry, and material science\cite{sethian_levelset}.

\begin{figure}[h]
	\includegraphics[width=3.33in]{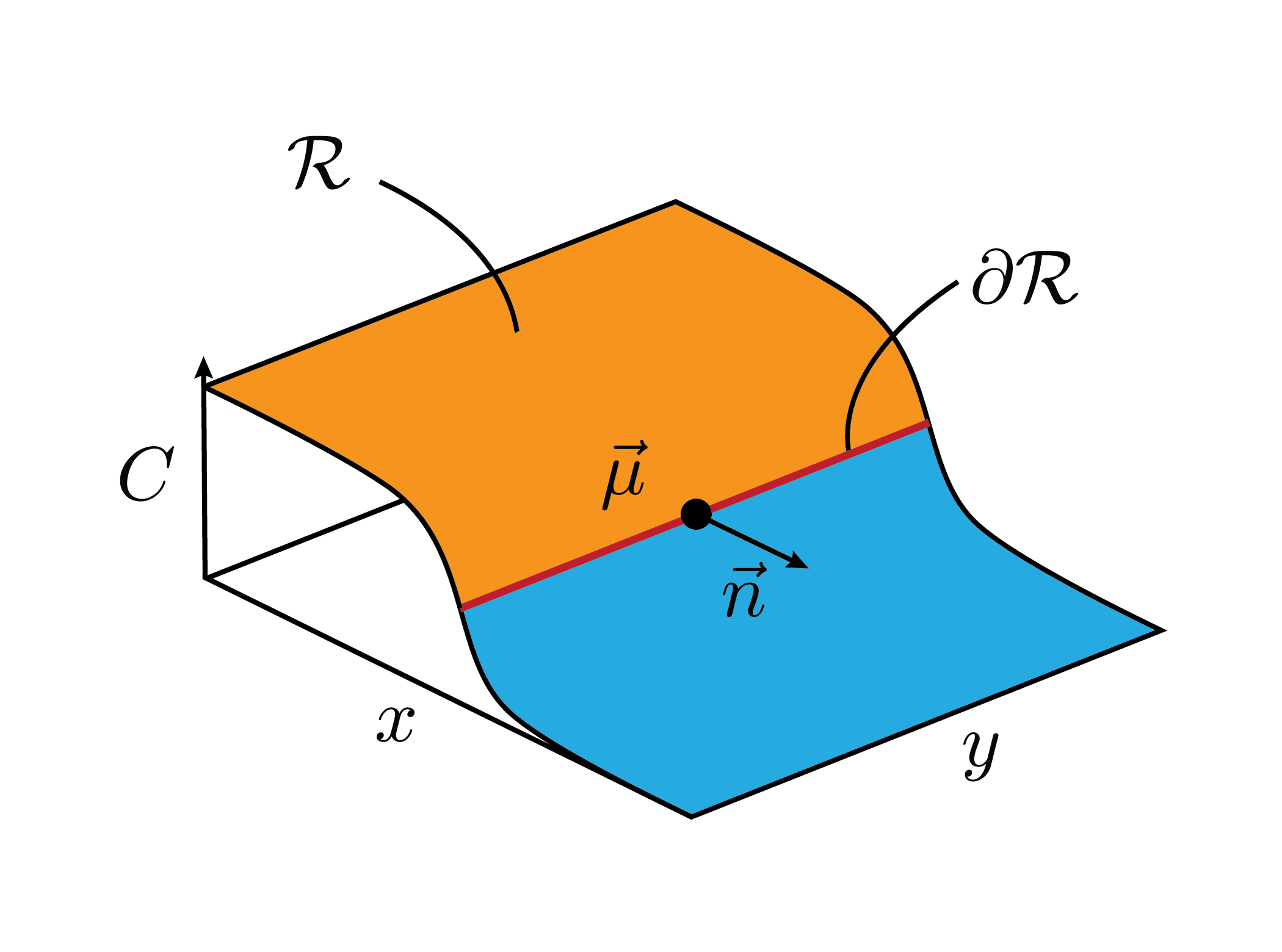}
	\caption{A graph of a three-dimensional planar counting function (equation \ref{eq-linear_cf}) projected into the x-y plane. The counting region $\mathcal{R}$ is indicated in orange and the boundary $\partial \mathcal{R}$ as a red line. The boundary is the plane normal to $\vec{n}$ that intersects the point $\vec{\mu}$ and is the space where the counting function value equals $1/2$. }
	\label{fig-linear_cf}
\end{figure}

\begin{figure}[h]
	\includegraphics[width=3.33in]{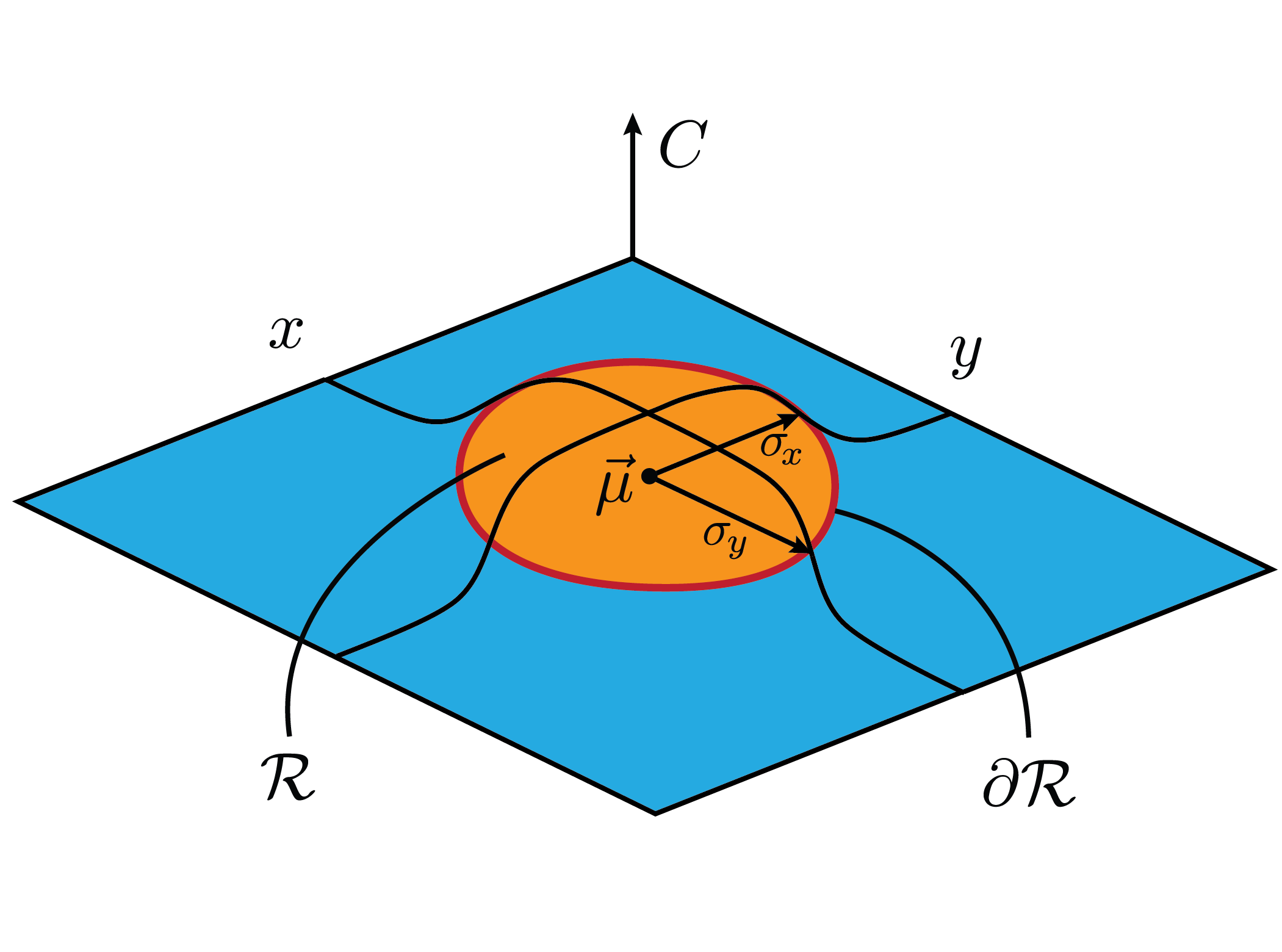}
	\caption{A graph of a three-dimensional ellipsoidal counting function projected into the x-y plane. The counting region $\mathcal{R}$ is indicated in orange and the boundary $\partial \mathcal{R}$ is indicated as a red line. The boundary is the ellipse centered at $\vec{\mu}$ with axis length and directions given by the eigenvalues and eigenvectors of $\mat{A}^{-1/2}$ respectively and the space where the counting function value equals $1/2$.  }
	\label{fig-quad_cf}
\end{figure}

With this analysis in place, we start by looking at switching surfaces that emerge from simple functional forms of $f$. When $f$ is a linear form:
\begin{equation} \label{eq-linear_cf}  f_l(\vec{r}) = -\vec{n}\cdot(\vec{r} - \vec{\mu}), \quad C_l = \frac{1}{1 + \exp\left( -\vec{n}\cdot(\vec{r} - \vec{\mu})\right)}\end{equation}
The switching surface is a plane centered at $\vec{\mu}$ and normal to $\vec{n}$. 
Likewise, when $f$ is a quadratic form:
\begin{align}
f_{q}(\vec{r}) &= (\vec{r} - \vec{\mu}) ^T \mat{A} (\vec{r} - \vec{\mu}) - K,  \notag \\
C_{q} &= \frac{1}{1 + \exp\left( K\left(\left(\vec{r} - \vec{\mu}\right)^T \mat{A} \left(\vec{r} - \vec{\mu} \right) - 1\right) \right)}
\label{eq-quad_cf}
\end{align}
the switching surface adopts a spherical, ellipsoid, paraboloid, or hyperboloid shape, depending on the sign-structure of the eigenvalues of $\mat{A}$, centered at the point $\vec{\mu}$. 
Labeled examples of each of these counting functions are illustrated in figures \ref{fig-linear_cf} and \ref{fig-quad_cf}, and we will often refer to counting functions by the shape of their switching surface.

\begin{figure}[h]
	\includegraphics[width=3.33in]{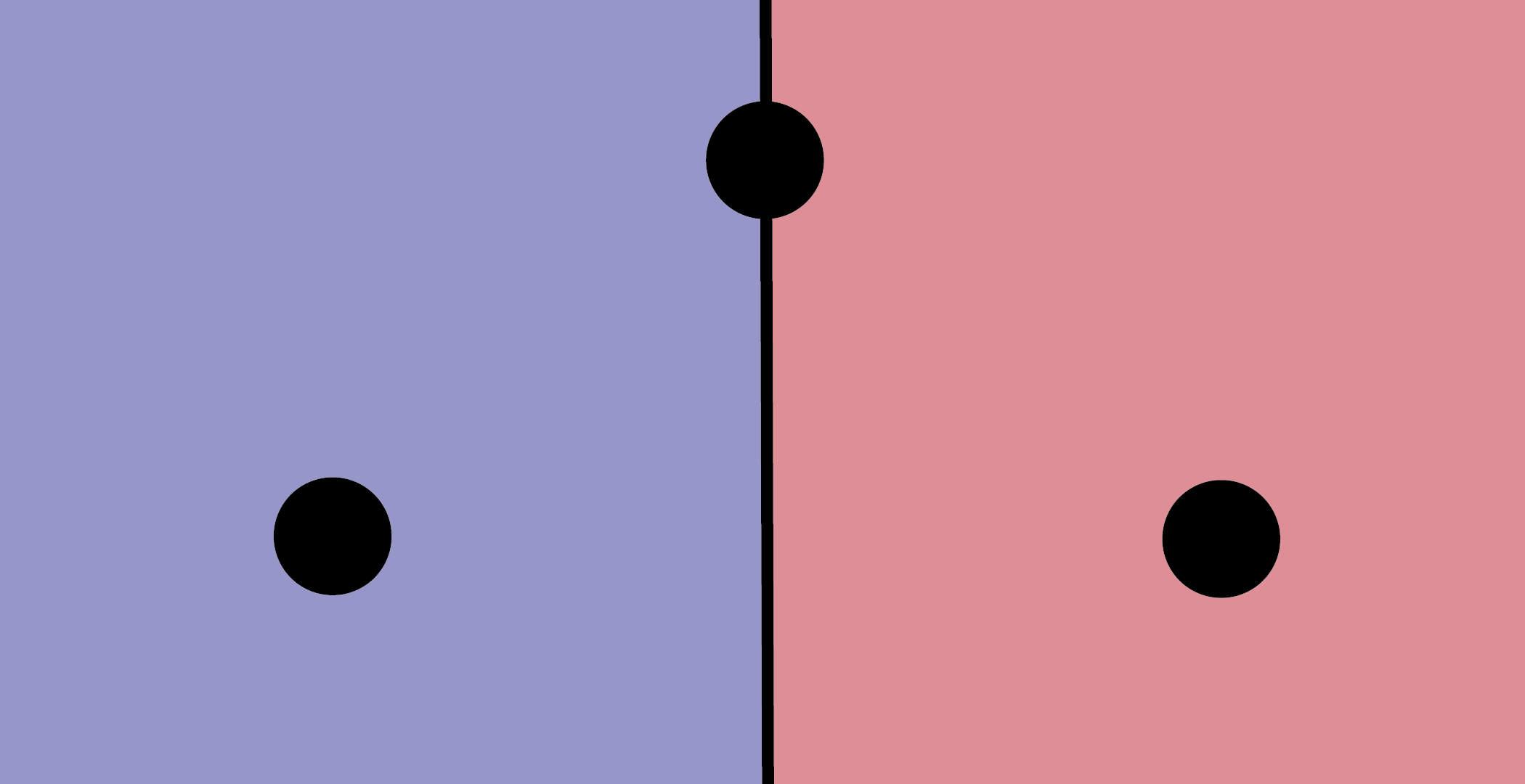}
	\caption{In a na\"ive generalization to a simple two-fragment system, 
    we place a pair of planar counting functions bisecting each bond axis back-to-back.
    Shown is an isosceles arrangement of atoms (black circles) with a pair of planar counting functions bisecting the base of the triangle.
    The counting regions are indicated by blue and red regions and the switching surface by the black line.
    The switching surface intersects the atom at the top vertex, and introduces a kinetic energy cost that limits particle-number projections between electronic populations of the atoms at the base.}
	\label{fig-planar_ke_bias}
\end{figure}

\begin{figure}[h]
	\includegraphics[width=2.00in]{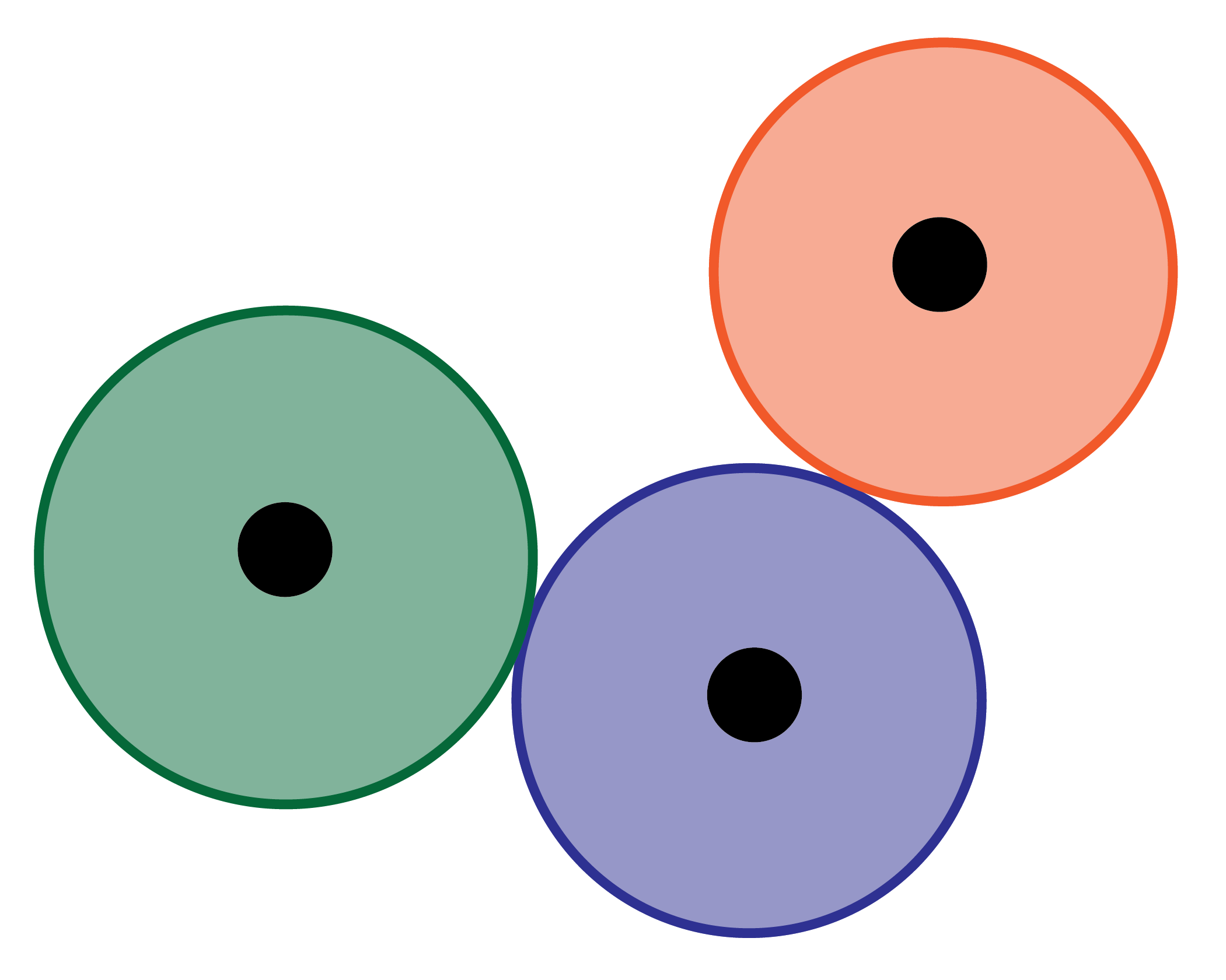}
	\caption{An illustration of a set of ellipsoidal regions (indicated by the shaded regions) centered at atomic positions indicated by the black circles. Spherically packing counting functions always produces an incomplete tiling of space, leaving significant curvature at the void boundary. Due to the curved boundary of the counting regions, the interface between neighboring counting regions is very narrow, making it impossible for particles to travel between them without encountering this void curvature and incurring a kinetic energy cost to prospective number projections.
    As the curvature from neighboring regions does not cancel out in simple linear combinations, attempting number projections between multi-atom fragments using linear combinations of counting functions will encounter this kinetic energy cost as well.
    }
	\label{fig-fd_ellipsoidal_problem}
\end{figure}

We have shown that a pair of properly placed planar counting functions equip the counting Jastrow with the ability to project out ionic terms in a molecular dissociation process\cite{vdg_ncjf}.
However, this relies on an \emph{ad hoc} placement of the counting functions across the bond-axis, and immediately runs into problems if na\"ively generalized.
The gradient and Laplacian of the counting Jastrow exponent $J_C$ attain extreme values near the switching surface, and when the switching surface overlaps with the wavefunction reference orbitals, these terms can easily heighten the kinetic energy of the wavefunction:
\begin{align}
& \braket{e^{J_C}\psi | \op{T} | e^{J_C} \psi} \notag \\
& \hspace{5mm}
= -\frac{1}{2}\int |e^{J_C}\psi|^2
\Big[
        \nabla^2 J_C
      + |\nabla J|^2
      +  \frac{2 \nabla J_C \cdot\nabla \psi}{\psi}
\notag \\
& \hspace{35mm}
      + \frac{\nabla^2\psi}{\psi}
\Big] d\ \{\vec{r}_i\}
\label{eq-kinetic_energy}
\end{align}
Kinetic energy introduced by these terms is not necessarily unphysical, as the electronic ground state performs a balancing act between kinetic and potential energy in which the counting Jastrow may freely participate.
However, simply bisecting every chemical bond with a pair of planar counting functions in even trivial chemical systems can accidentally place the switching surface on top of a distant atom, as shown in figure \ref{fig-planar_ke_bias}.
In many of these cases, it is easy to imagine modifying the counting region boundary to just avoid the distant nuclear center while making only a small change to the counting region itself, ultimately lowering the kinetic energy of the wavefunction with little to no tradeoff in potential energy.
As a result, the number projecting action of the counting Jastrow is limited by the uncontrolled overlap between these planar switching surfaces and distant reference orbitals in this hypothetical scheme.

\begin{figure}[t]
    \centering
	\includegraphics[width=3.33in]{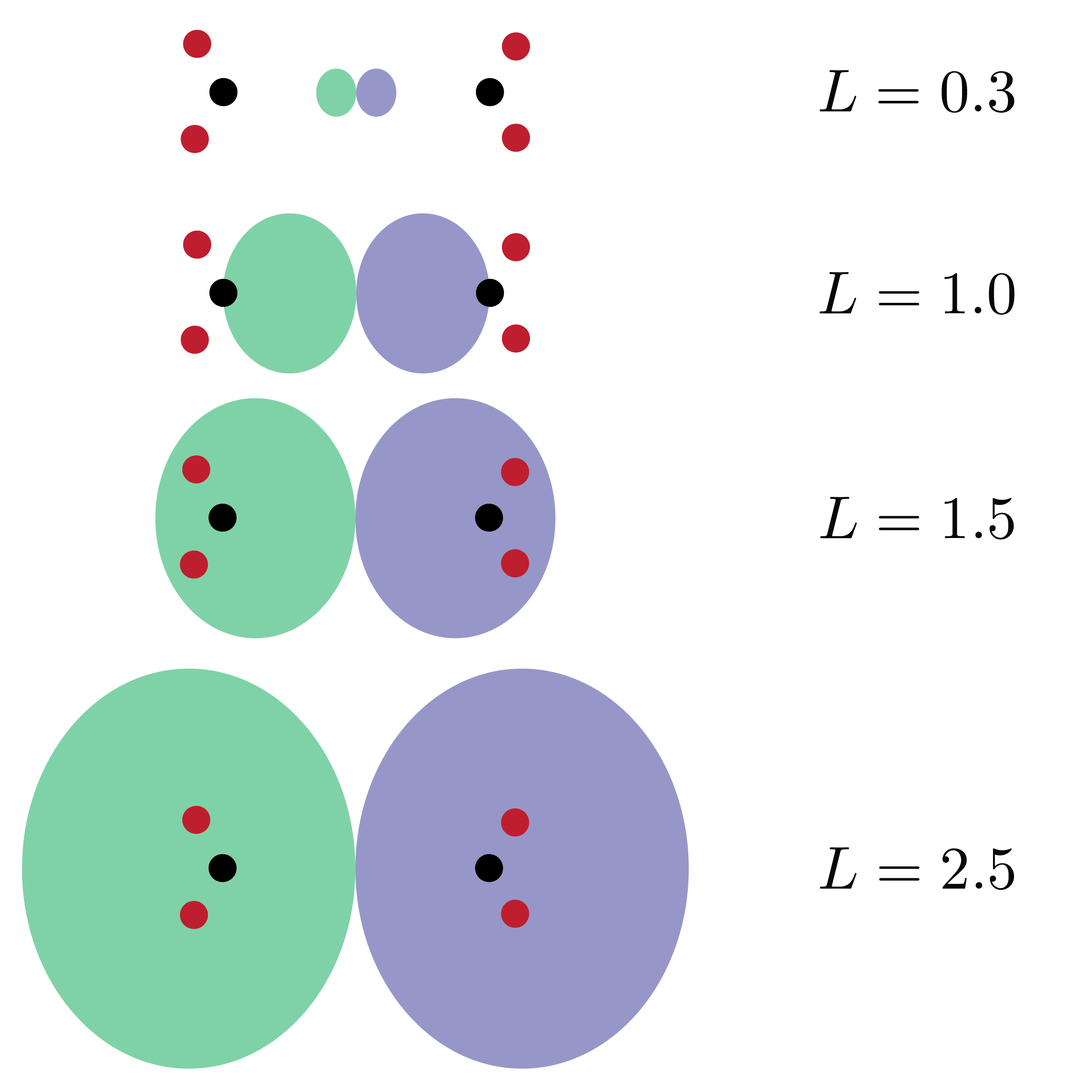}
	\caption{Schematic of the counting regions as a function of the scanning coordinate used in figure \ref{fig-ethene_overlap_energy}.
    Single-point VMC energy calculations were performed for ethene with a distance of 5 $\mbox{\AA}$ between carbon atoms as a function of the counting region scale factor L, indicated in the figure, where the pair of ellipsoidal counting regions were set to touch at the bond midpoint.
    The number projection was performed using $\mat{F}$ matrix parameters optimal for a pair of bond-bisecting planar counting functions.}
	\label{fig-ethene_overlap_schematic}
\end{figure}

\begin{figure}[t]
    \centering
	\includegraphics[width=3.33in]{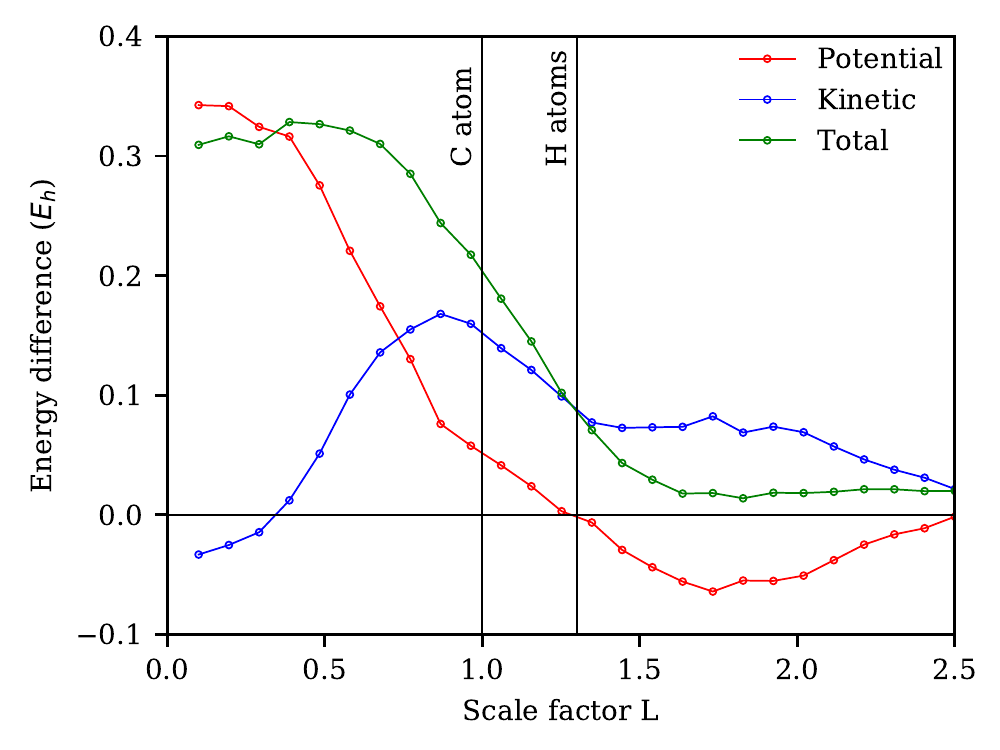}
	\caption{Plot of potential, kinetic, and total VMC energies for ethene using a 
    CJS wavefunction with a counting function basis of scaled ellipses, shown in figure \ref{fig-ethene_overlap_schematic} relative to VMC energies of a CJS wavefunction using a pair of bond-bisecting counting functions.
    Scale factor values that correspond to an intersection of the ellipsoidal switching surfaces with atomic centers are indicated by the labeled vertical lines.
    As the ellipsoidal regions completely encompass each fragment, the projecting action of the ellipsoid counting functions approaches that of planar counting functions. 
    At no point do the atom-centered ellipsoidal counting functions improve the total variational energy of the CJS wavefunction relative to the planar CJS benchmark indicated by the horizontal line.}
	\label{fig-ethene_overlap_energy}
\end{figure}

Thinking instead that finite, enclosed boundaries won't encounter this problem, we attempted to generalize this basis by placing a single spherical counting function around each atom, like in figure \ref{fig-fd_ellipsoidal_problem}.
However, as shown in figures \ref{fig-ethene_overlap_schematic} and \ref{fig-ethene_overlap_energy}, when stretching ethene symmetrically across the central double bond, we were unable to find an ellipsoidal counting function basis which improved the variational energy of a CJS wavefunction relative to one that used a pair of planes to bisect the chemical bond.
Of course, the fundamental mathematical reason for this failure is the same: the switching surface of the ellipsoidal regions overlap with reference orbitals and limit the scope of variationally favorable number projections.
In this case, the overlap originates from the incomplete spatial packing of the spherical switching surfaces which produce void regions that permeate a significant fraction of space, greatly restricting number projections from the counting Jastrow no matter how these ellipsoidal regions are placed.

\subsection{Normalized Counting Functions}
\label{sec-normalized_counting_functions}
The collective pathologies of these schemes suggest that an improved counting function basis will have the following attributes:
\begin{enumerate}
\item \emph{Localizability}: regions permit finite boundaries.
\item \emph{Completeness}: regions completely tile space.
\item \emph{Clean Additivity}: summation seamlessly combines counting functions.
\end{enumerate}
In this study, we look at counting functions of the form:
\begin{equation} \label{eq-normgauss} C_I(\vec{r}) = \frac{g_I(\vec{r})}{\sum\limits_{J} g_J(\vec{r})} \end{equation}
with the three-dimensional gaussian functions $g_I$:
\begin{align}
g_I(\vec{r}) &= \exp\left( \vec{r}^T \mat{A}_I \vec{r} - 2\vec{B}_I^T \vec{r} + K_I\right) \notag \\
             &= \exp\left( (\vec{r} - \vec{\mu}_I)^T \mat{A}_I (\vec{r} - \vec{\mu}_I) + \tilde{K}_I  \right)
\end{align}
which fulfills all three of these conditions.
Note that normalized gaussian functions like this are widely used in statistical classification algorithms such as quadratic discriminant analysis\cite{mclachlan_discriminant_analysis} and gaussian mixture models\cite{mclachlan_mixture_models}, which should not be too surprising, as the fundamental goal of our counting functions is to classify particles according to their location.
We will refer to counting functions of this form as normalized counting functions built from the anchor gaussians $g_I$. 

The relationship between the gaussian parameters $\{\mat{A}_I, \vec{B}_I, K_I\}$ and the counting regions they produce is not immediately clear, but upon further investigation, we find that the behavior of a normalized counting function can be modeled by what we will call a `pair counting function' throughout most points in space.
Pair counting functions are those formed from distinct pairs of anchor gaussians and are mathematically identical to the counting functions discussed in equation \ref{eq-quad_cf}:
\begin{equation}\label{eq-pair_cf} C_{I/J}(\vec{r}) = \frac{g_I(\vec{r})}{g_I(\vec{r}) + g_J(\vec{r})} = \frac{1}{1+g_J(\vec{r})/g_I(\vec{r})}\end{equation}
where the first element of the compound index $I/J$ indicates the index of the gaussian in the numerator.
These pair counting functions approximate normalized counting functions for most of space in the sense that, for $I \ne J$:
\begin{enumerate}
\item $C_I(\vec{r}) \approx C_{I/J}(\vec{r})$ when $g_I(\vec{r})$ and $g_J(\vec{r})$ are the largest two anchor gaussians.
\item $C_{I}(\vec{r}) < C_{I/J}(\vec{r})$ everywhere.
\end{enumerate}
The space where the first property does not hold for any pair of anchor gaussian is limited to neighborhoods about space where three anchor gaussians are equal:
\begin{equation} \label{eq-cf_edge} \{\vec{r}: g_I(\vec{r}) = g_J(\vec{r}) = g_K(\vec{r}) \}\end{equation}
As this set is defined by two independent constraints, it defines a one-dimensional path in space.
More precisely, since normalized counting functions are continuous, this condition holds, to within a uniform convergence factor, except at a small neighborhood about these paths. 
The size of these neighborhoods shrink as we increase the gradient of the counting function at the switching surface -- which is easily done by uniformly scaling the variance of all anchor gaussians -- and vanishes as the gradient diverges.
These pair counting functions are still good approximations to the boundary even at modest gradient values, and we can treat the switching surface of these normalized counting functions as a patchwork of quadratic surfaces formed by the pair counting functions, joined together at edges given by the paths in equation \ref{eq-cf_edge}.

We define the counting region $\mathcal{R}_I$ as the region in which the normalized counting function is greater than one-half:
\begin{equation} \mathcal{R}_I = \{\vec{r}: C_I(\vec{r}) > 1/2\}\end{equation}
since the second property holds for all $J\ne I$, it follows that:
\begin{equation} C_I(\vec{r}) < \min\limits_{J\ne I} C_{I/J}(\vec{r}) \end{equation}
so in order for a point to be contained in $\mathcal{R}_I$, it must also be contained in every pair counting region $\mathcal{R}_{I/J}$.
As the region boundaries coincide with the pair region boundaries to within the aforementioned convergence factor, we conclude that $\mathcal{R}_I$ is well approximated as the intersection of the pair counting regions:
\begin{equation} \mathcal{R}_I \approx \bigcap\limits_{I\ne J} \mathcal{R}_{I/J}\end{equation}
to within this convergence factor.
As a result, the counting regions of these normalized counting functions can be understood in terms of these pair counting functions, and we can use them study their boundaries facet by facet.

When this analysis is accurate, which it is nearly everywhere in space, these normalized counting functions are localizable, complete, and cleanly additive, as defined above.
First, these normalized counting functions are localizable simply because they can be bounded into finite domains, either by a single ellipsoidal boundary or by multiple planar boundaries.
Secondly, due to the normalization condition, the tiling is complete, as the entire set of counting functions account completely for each particle at every point in space:
\begin{equation}\label{eq-pointwise_additivity_start} \sum\limits_{I} C_I(\vec{r}_i) = 1 \end{equation}
Finally, these are cleanly additive, since the pair counting functions along the shared boundary between adjacent regions sum to one:
\begin{equation} C_{I/J}(\vec{r}) = 1 - C_{J/I}(\vec{r})\end{equation}
and the contracted counting function, given by the sum:
\begin{equation}C_{I+J} = C_I + C_J\end{equation}
has no seam along the shared switching surface and acts as though its counting region is the union of the components' counting regions (to within some small convergence factor):
\begin{equation} \label{eq-pointwise_additivity_end}  \mathcal{R}_{I+J} \approx \mathcal{R}_I \bigcup \mathcal{R}_J \end{equation}
These contractions can be exactly represented in the linear Jastrow coefficients $\mat{F}$ and $\vec{G}$ featured in equation \ref{eq-cf_def}, allowing variational methods to freely adjust them to form clean region combinations --  a feat that was not always possible in the previous basis formulation.
Thus, this normalized-gaussian functional form is able to perform number projections between composite regions without creating curvature in the interior of the contracted counting functions and unduly raising the kinetic energy of the wavefunction in the process.

Anchor gaussians are so named because their mean positions serve as central anchoring points for the normalized counting functions, and roughly embody their geometric centers, as will become more evident in the partitioning schemes that follow.
The normalization condition in equation \ref{eq-normgauss} used to generate these counting regions is at the heart of their convenient formal properties that allow them to automatically form regions with highly flexible and well-characterized quadratic boundaries. 
However, this normalization condition induces redundant wavefunction representations in the Jastrow parameter space, which manifest linear dependencies in the wavefunction tangent space and cause ill-conditioned numerics in variational optimization algorithms.
In Appendix \ref{app-linear_dependencies}, we trace the cause of these issues and describe how to remove this undesired behavior in the context of the Linear Method.

\section{Partitioning Schemes}
\subsection{Classical Voronoi Partitioning}
\label{sec-classical_voronoi}

In a classical Voronoi tessellation\cite{aurenhammer2013voronoi}, space is divided into disjoint regions according to its distance from a set of Voronoi points $\{\vec{v}_I\}$.
Each Voronoi point maps to a subset of space $\mathcal{V}_I$ which is defined as the region that is closer to that Voronoi point than any other:
\begin{equation}\label{eq-voronoi_condition} \mathcal{V}_I = \left\{ \vec{r}: \| \vec{r} - \vec{v}_I \| < \|\vec{r} - \vec{v}_J\|, \text{ for all } I \ne J \right\} \end{equation}
Each of these Voronoi regions can be written as the intersection of Voronoi regions built solely from pairs of Voronoi points:
\begin{equation} \mathcal{V}_I = \bigcap\limits_{J\ne I} \mathcal{V}_{I/J}, \quad \mathcal{V}_{I/J} = \{ \vec{r}: \|\vec{r} - \vec{v}_I\| < \| \vec{r} - \vec{v}_J \| \}\end{equation}
The normalized counting functions described in section \ref{sec-normalized_counting_functions} can be parameterized to divide space in a nearly identical way.

When the quadratic parameters $\mat{A}_I$ for each anchor gaussian in a set of normalized counting functions are equivalent, the counting regions are equivalent to (within some convergence factor) Voronoi regions generated from the mean gaussian positions.
For simplicity, we will look at the case where these matrices are all equal to some scalar multiple of the identity matrix:
\begin{equation} \mat{A}_I \equiv \alpha \mat{I}\end{equation}
With this restriction on $\mat{A}_I$, these pair counting functions can be manipulated to adopt the general form of a planar counting function given in equation \ref{eq-linear_cf}:
\begin{equation} C_{I/J} = \frac{1}{1 + \exp\left[ -4 \alpha \left( \vec{r} - \left( \frac{ \vec{\mu}_I + \vec{\mu}_J }{2} \right) \right) \cdot  \left( \frac{\vec{\mu}_J - \vec{\mu}_I}{2} \right) \right]} \end{equation}
with:
\begin{equation}\vec{\mu} = \frac{\vec{\mu}_J + \vec{\mu}_I}{2}, \quad  \vec{n} = \frac{\vec{\mu}_J - \vec{\mu}_I}{2}\end{equation}
As a result, the pair counting region $\mathcal{R}_{I/J}$ is identical to the Voronoi regions $\mathcal{V}_{I/J}$ defined above, and following section \ref{sec-normalized_counting_functions}, the counting regions can be written as the intersection of the pair counting regions (to within some convergence factor):
\begin{equation} \mathcal{R}_I \approx \bigcap \limits_{J \ne I} \mathcal{R}_{I/J}\end{equation}
which echoes the Voronoi condition given in equation \ref{eq-voronoi_condition}.

One immediate application of this scheme is to place anchor gaussians such that their means coincide with each atomic center, which generates a set of atom-centered Voronoi counting regions.
The sum of normalized counting function values over electronic coordinates generated in this way roughly corresponds to an atomic population analysis performed on an individual particle configuration.
As a result, these partitions are able to identify and remove interatomic ionic terms such as those that appear in dissociation processes, with the added benefit of a systematic and automated generation scheme, unlike the simpler counting function basis used previously.
\subsection{Spherical Voronoi Partitioning}
\label{sec-spherical_voronoi}

Spherical coordinates are a natural choice in chemical systems due to the strong isotropic, attraction between electrons and nuclei.
Likewise, counting regions that partition space into spherical sectors organically describe correlations between electrons near the same atomic center.
One might imagine approximating this partition using the scheme described in section \ref{sec-classical_voronoi} by surrounding an atomic center with a set of anchor gaussians placed at the vertices of a closed polyhedron.
An increasingly fine angular mesh of the resulting flat-faced Voronoi regions converge to the desired spherical partition, which could be recombined to form any angular shape using the property of pointwise additivity described in section \ref{sec-normalized_counting_functions}.
Though technically exact in the infinite limit, rough constructions like this can  be unwieldy and imprecise, increasing either computational cost or introducing an unknown convergence error in cases where exact spherical boundaries would be a better choice.

\begin{figure}
    \centering
	\includegraphics[width=3.33in]{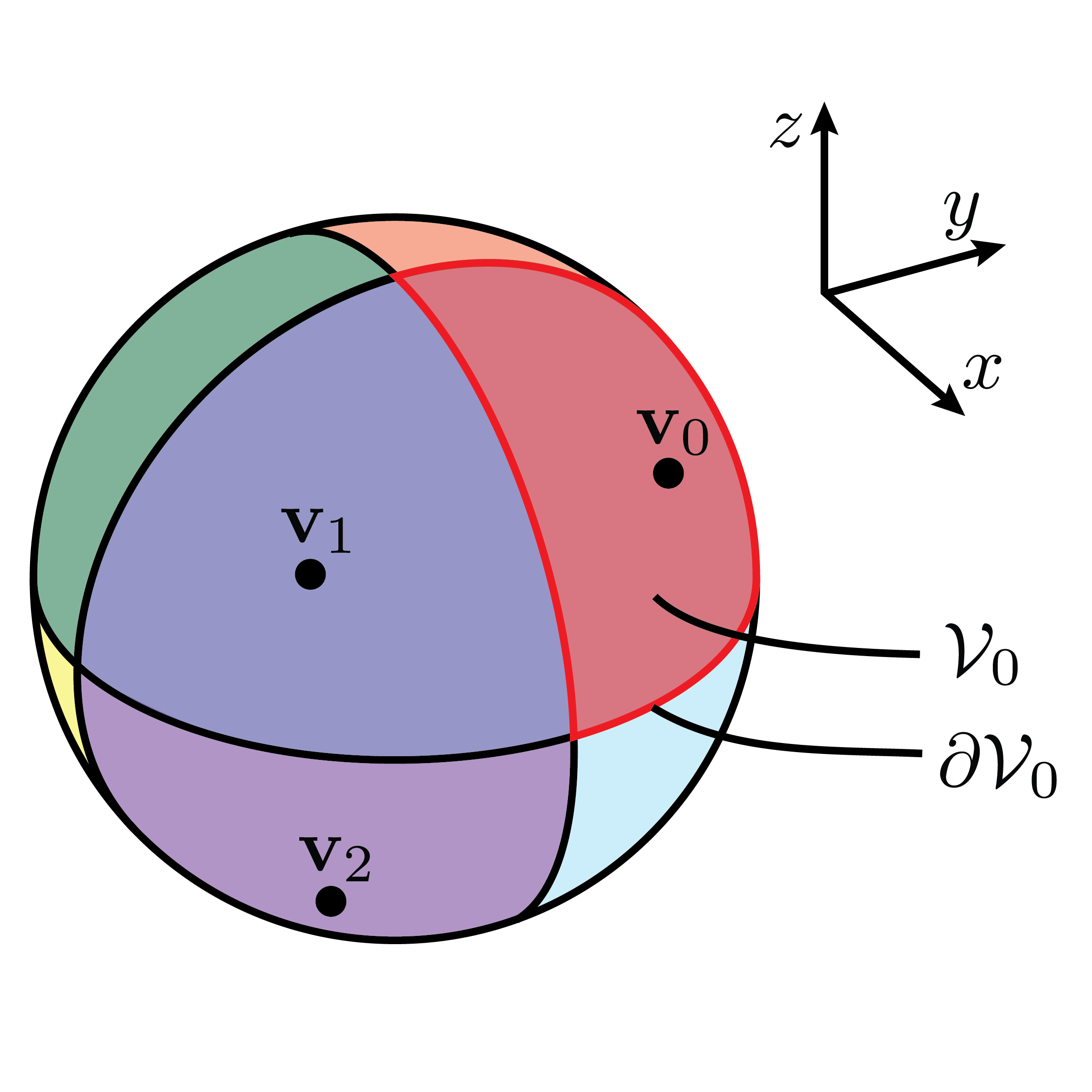}
	\caption{A spherical Voronoi diagram, in which the surface of a sphere is partitioned into angular sections based on the surface arc-length from a set of Voronoi points, indicated by labeled black circles.
    A  planar division  between these points in three dimensions is sufficient to produce these boundaries, and figure \ref{fig-spherical_voronoi_extrude} shows how these sectors can be further subdivided using spherical boundaries.}
	\label{fig-spherical_voronoi}
\end{figure}

As discussed in the previous section, classical Voronoi tessellations are a natural way to partition space based on distances from a set of predetermined points.
This Voronoi partitioning scheme may be easily adapted to partition a spherical surface in a directly analogous way, as depicted in figure \ref{fig-spherical_voronoi}.
Formally, these partitions are defined by restricting our attention to the surface of a sphere and subdividing the surface regions according to the Voronoi condition (equation \ref{eq-voronoi_condition}) with the distance measured by great-arcs on the spherical surface\cite{aurenhammer2013voronoi}.
Fortunately, we can generate this partition using the classical Voronoi scheme described in the previous section by placing each anchor gaussian on the surface of this sphere, as the planar surfaces intersect the center of the sphere and produce the correct geodesic boundaries.
As this generates only an angular partition, we will extend this scheme into three dimensions by further subdividing this partition into radial shells.
Luckily, the normalized counting functions described in section \ref{sec-normalized_counting_functions} can easily accommodate the curved radial boundaries through the quadratic parameters $\mat{A}_I$ present in the anchor gaussians.

\begin{figure}
    \centering
	\includegraphics[width=3.33in]{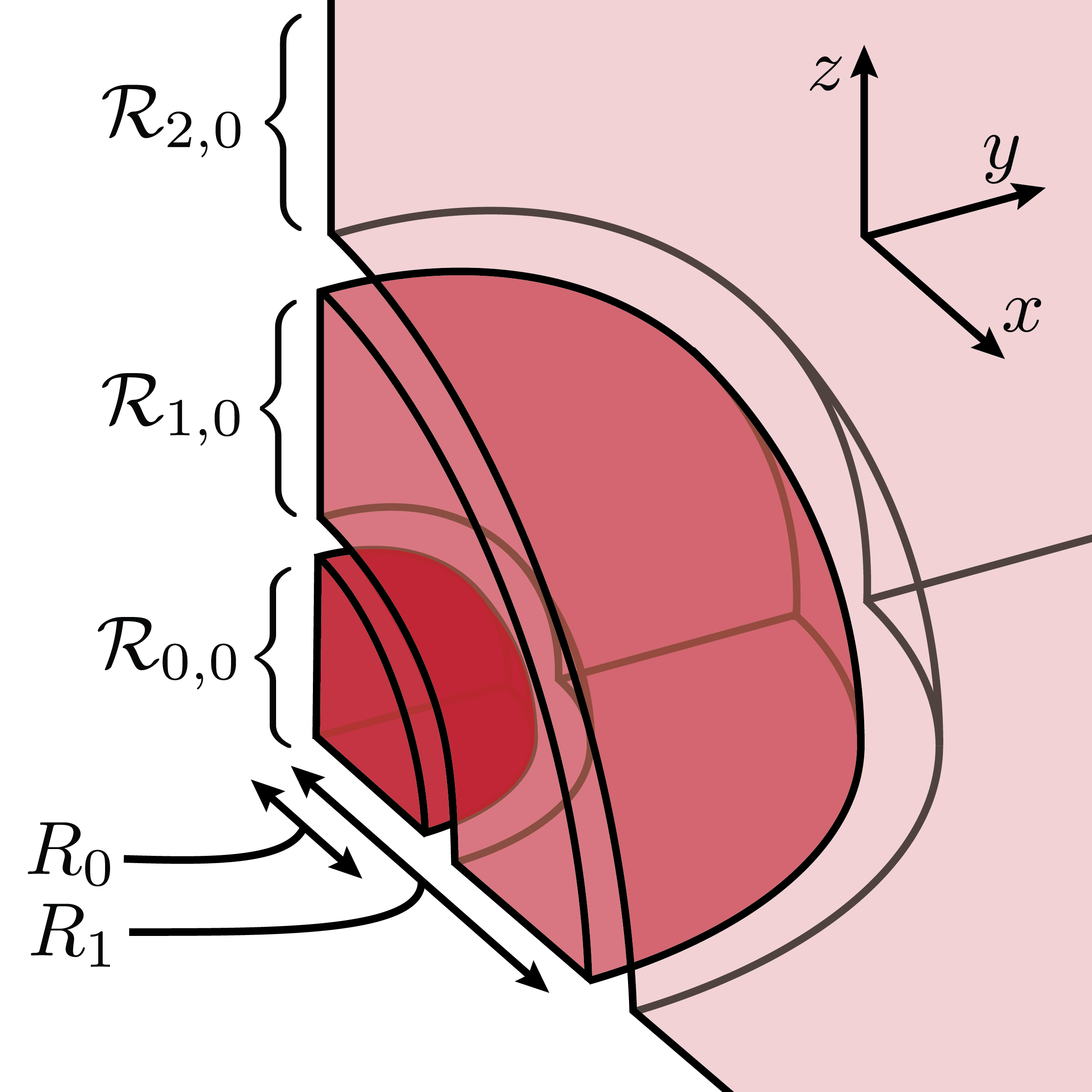}
	\caption{The angular sector in figure \ref{fig-spherical_voronoi} subdivided into three radial shells with radial boundaries at $R_0$ and $R_1$. Gaps between regions are shown to highlight the location of radial boundaries. 
    In this partitioning scheme, the other spherical regions in figure \ref{fig-spherical_voronoi} are radially subdivided in exactly the same way.}
	\label{fig-spherical_voronoi_extrude}
\end{figure}

An algorithm to generate these spherical Voronoi partitions complete with explicit formulae for anchor gaussian parameters is given in appendix \ref{app-spherical_voronoi}, which we summarize here.
First, we place $N_{\Omega}$ anchor gaussians with identical quadratic parameters $\mat{A}_I$ whose means $\vec{\mu}_I$ lie on the surface of a sphere, following the Voronoi scheme in section \ref{sec-classical_voronoi}.
When these anchor gaussians are normalized to create normalized counting functions, the counting regions take on the shape of angular sectors with boundaries that coincide with the spherical Voronoi diagram generated from the means of the anchor gaussians shown in figure \ref{fig-spherical_voronoi}.
Second, we choose a set of $N_R$ radii along which we will subdivide each of these angular sectors starting from the innermost shell and successively create spherical divisions between adjacent shells.
To add a single radial shell of counting regions, for each anchor gaussian that lay on the current outermost shell, we place another at the same angular coordinates with parameters chosen such that the pair counting functions made from these two anchor gaussians have switching surfaces on a sphere at the prescribed radius, midway between the two anchor gaussians.
This generates a total of $N_{\Omega}\cdot(N_{R} + 1)$ counting regions where the curvature of the radial partition is not limited by the number of angular divisions.
As depicted in figure \ref{fig-spherical_voronoi_extrude}, the final angular sectors subdivided into radial shells can be visualized as radial extrusions of the initial spherical Voronoi partition.

\subsection{Region Composition}
\label{sec-region_composition}
\begin{figure}
    \centering
	\includegraphics[width=3.33in]{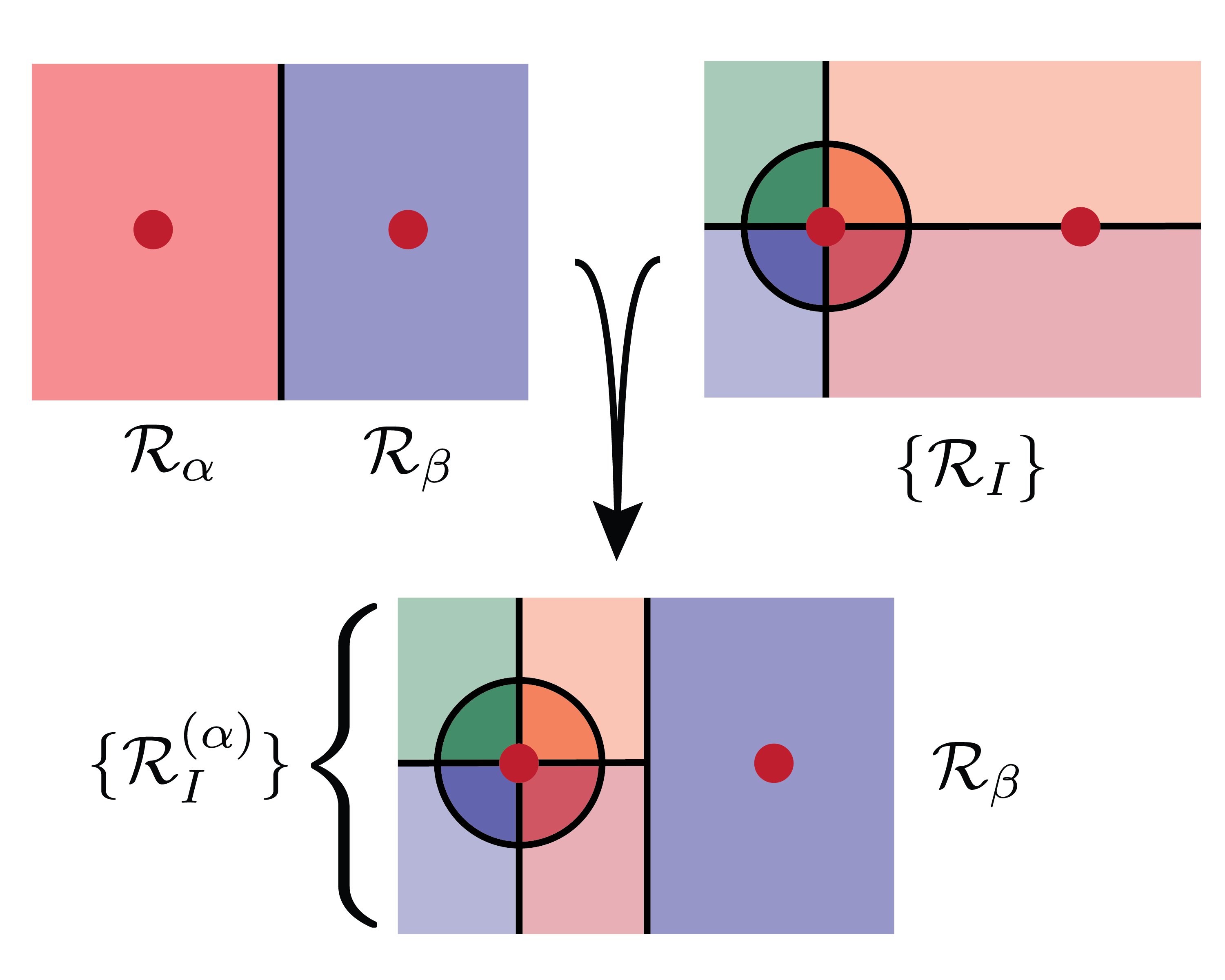}
	\caption{Schematic depicting a composition which subdivides a single atom-centered Voronoi region $\mathcal{R}_{\alpha}$ into a set of regions $\{ \mathcal{R}_I^{(\alpha)}\}$ that best match the given partition $\{ \mathcal{R}_I\}$.
    Counting regions are indicated by the shaded areas, atomic positions are indicated by the red circles, and switching surfaces are indicated by black lines.
    To divide space according to both partitions simultaneously, we must determine the $g_I^{(\alpha)}$ that match the conditions in equations \ref{eq-cond1} and \ref{eq-cond2}.
    }
	\label{fig-composition_labelled}
\end{figure}

Both of these schemes are useful and complete and each is most appropriate at describing correlations at different scales in the context of molecular electronic structure. 
The classical Voronoi partitioning scheme described in section \ref{sec-classical_voronoi} divides space into a set of atom-centered Voronoi cells, naturally correlating electronic populations between different atoms. 
The spherical partitioning scheme in section \ref{sec-spherical_voronoi} instead divides space into a set of spherical sections and is most suited to capture correlations within a single atomic shell.
As we look to apply these counting Jastrows in more complex systems, combining these two approaches within a single normalized counting function basis appears a promising way to describe correlations at both of these scales simultaneously.
In addition, subdividing existing partitions while retaining the existing divisions allows us to approach the basis set limit for these counting functions in a systematic and chemically sensible way.
In the following discussion, we will consider a rough partition consisting of two regions, $\mathcal{R}_{\alpha}$ and $\mathcal{R}_{\beta}$, and a fine-grained partition $\{\mathcal{R}_I\}$ which will be used to subdivide $\mathcal{R}_{\alpha}$, as depicted in figure \ref{fig-composition_labelled}.
Our goal is to combine the two partitions into a single normalized counting function basis in a way that best preserves their boundaries simultaneously. 

The property of clean additivity discussed in section \ref{sec-normalized_counting_functions} states that simple sums of adjacent counting functions act much like a single counting function over their combined region.
This suggests a natural condition when splitting single counting function ($C_{\alpha}$) into a set of pieces ($C_I^{(\alpha)}$):
\begin{equation} C_{\alpha}(\vec{r}) = \sum\limits_I C_{I}^{(\alpha)}(\vec{r})\end{equation}
In order to retain the formal properties outlined in section \ref{sec-normalized_counting_functions}, counting functions split in this way must be built from a single set of anchor gaussians.
Our task is to find a set of $g_I^{(\alpha)}(\vec{r})$ that both reproduces the divisions of the fine-grained partition internal to $\mathcal{R}_{\alpha}$:
\begin{equation} \label{eq-cond1}C_I^{(\alpha)}(\vec{r}) = \frac{g_I^{(\alpha)}(\vec{r})}{\sum\limits_I g_I^{(\alpha)}(\vec{r})  + g_{\beta}(\vec{r}) } = \frac{g_I(\vec{r})}{\sum\limits_{J} g_J(\vec{r})}\end{equation}
and contracts to reproduce the counting function they replace:
\begin{equation} \label{eq-cond2}C_{\alpha}(\vec{r}) = \frac{g_{\alpha}(\vec{r})}{g_{\alpha}(\vec{r}) + g_{\beta}(\vec{r})} = \frac{\sum\limits_J g_J^{(\alpha)}(\vec{r})}{\sum\limits_{J} g_J^{(\alpha)}(\vec{r}) + g_{\beta}(\vec{r}) }    \end{equation}

Since the degrees of freedom we use to fulfill these conditions  are not immediately apparent, producing this subdivision may seem either trivial or impossible.
A full explanation of how this can be done is given in appendix \ref{app-region_composition}, which we will summarize here by stating that we can exactly fulfill these conditions at a single point to second-order in a local Taylor expansion.
We will use this scheme to subdivide two atom-centered Voronoi regions into spherical sections, and in this paper we choose to satisfy these conditions at the midpoint of the chemical bond.
As the composite set of counting functions divides space into both atom-centered Voronoi regions and spherical subregions simultaneously, it strictly improves the granularity of the counting function basis.
This composition scheme thus sets up a basis to capture population correlations between subregions while retaining the ability to enact number projections between distinct molecular fragments.

\section{Results}
\label{sec-results}
Restricted Hartree-Fock (RHF), open-shell Hartree Fock (ROHF), second-order M{\o}ller-Plesset perturbation theory (MP2), coupled-cluster singles and doubles (CCSD) with perturbative triples (CCSD(T)), complete active space self-consistent field (CASSCF), and multireference configuration interaction singles and doubles with Davidson correction (MRCI+Q) calculations were performed using GAMESS \cite{gamess1,gamess2} and Molpro\cite{MOLPRO-WIREs,MOLPRO, molpro_integrals,molpro_ccsd_mp2, molpro_uccsd, molpro_ccsdt, molpro_mcscf1, molpro_mcscf2, molpro_mrci1, molpro_mrci2, molpro_mrci3}.
Due to the steep scaling of DMC calculations with atomic number\cite{dmc_core_scaling}, the helium core of carbon and oxygen and the neon core of calcium were replaced by energy-consistent pseudopotentials found in the Stuttgart library\cite{carbon_oxygen_ecp,calcium_ecp} in all calculations.
Variational Monte Carlo (VMC) and diffusion Monte Carlo (DMC) calculations\cite{qmc_review} were performed in a modified version of QMCPACK \cite{qmcpack} and the linear method\cite{linearmethod1,linearmethod2} was employed to minimize VMC energies.

We will indicate the functional form of wavefunction ansatzes according to the Jastrow factor exponents in table \ref{tab-jastrow_exponent} and acronyms in table \ref{tab-ansatz_acronyms}. 
The CJS exponent was initially set to zero and each TJS exponent was initially set to a minimal 10-parameter cubic basis spline that fulfilled both electron-nuclear and electron-electron cusp conditions.
Unless otherwise indicated, orbital optimizations were performed after pre-optimizing the Jastrow variables, and all variables were optimized together during orbital optimizations.

\begin{table}
  \caption{Jastrow exponent functional forms.}
  \begin{tabular}{c|c|c}
    Name & Symbol & Functional Form \\
    \hline
    One-Body Jastrow & $J_1$ & $\sum\limits_{i\alpha} u_1(r_{i\alpha}) $ \\
    Two-Body Jastrow & $J_2$ & $\sum\limits_{i\ne j} u_2(r_{ij}) $ \\
    Orbital rotation & $\op{X}$ & $\sum\limits_{ab} X_{ab} \op{a}_{a}^{\dagger} \op{a}_{b}, \quad \mat{X}^{\dagger} = 	-\mat{X}$ \\
    Counting Jastrow & $J_{NC}$ & $\sum\limits_{ijIJ} F_{IJ} C_{I}(\vec{r}_i) C_{J}(\vec{r}_j)$ \\
  \end{tabular}
  \label{tab-jastrow_exponent}
\end{table}

\begin{table}
  \caption{Ansatz acronyms and their corresponding functional forms. The reference wavefunction $\ket{\Psi_{HF}}$ is a single-determinant spin-restricted Hartree-Fock wavefunction.}
  \begin{tabular}{c|r}
   Acronym & Functional Form \\
    \hline
    TJS & $e^{J_1}e^{J_2}\ket{\Psi_{HF}}$ \\
    CJS & $e^{J_{C}} \ket{\Psi_{HF}}$ \\
    CTJS & $e^{J_1}e^{J_2}e^{J_{NC}} \ket{\Psi_{HF}}$\\
    TJS-oo & $e^{J_1}e^{J_2}e^{\hat{X}}\ket{\Psi_{HF}}$ \\
    CJS-oo  & $e^{J_{C}}e^{\hat{X}}\ket{\Psi_{HF}}$ \\
    CTJS-oo  & $e^{J_{C}}e^{J_1}e^{J_2}e^{\hat{X}} \ket{\Psi_{HF}}$\\
  \end{tabular}
  \label{tab-ansatz_acronyms}
\end{table}

\subsection{Ethene}
\label{sec-ethene}
\begin{figure}
    \centering
	\includegraphics[width=3.33in]{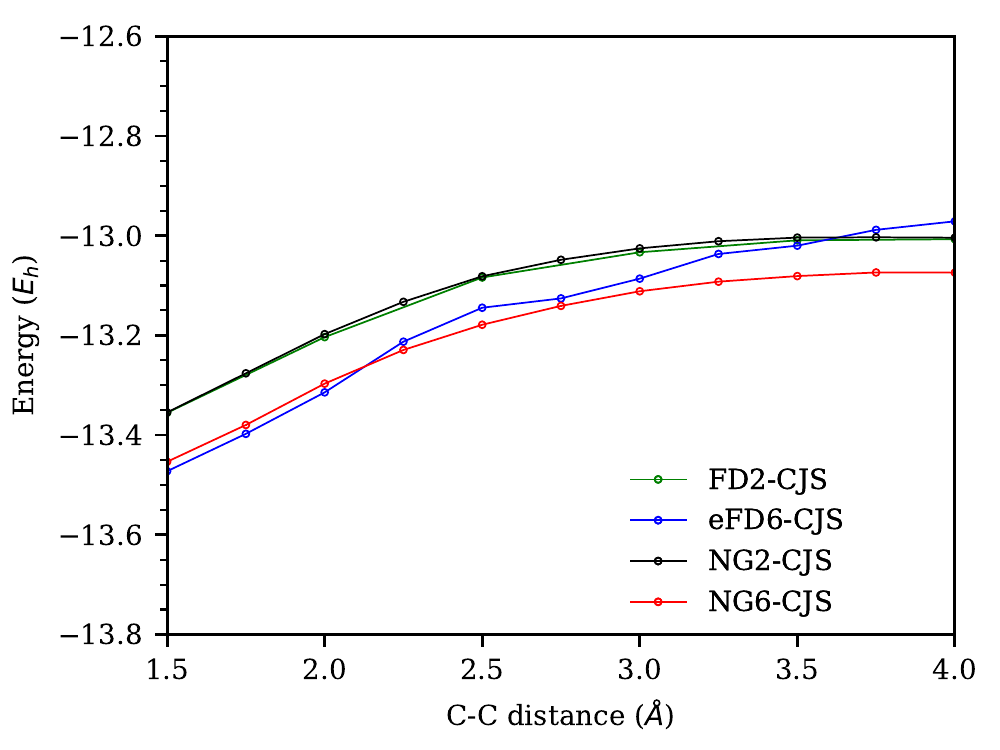}
	\caption{VMC energy as a function of ethene bondlength of CJS wavefunctions with a symmetry-adapted RHF reference using the counting function basis depicted in figures \ref{fig-c2h4_2ng_basis}, \ref{fig-c2h4_6ng_basis}, and \ref{fig-c2h4_e6fd_basis}. Fermi-Dirac style (FD) linear counting functions (equation \ref{eq-linear_cf}) are explicitly sigmoidal counting functions which are equivalent to a pair of carbon-centered normalized gaussians (NG2). Subdividing these counting functions into a total of six atom-centered Voronoi cells (NG6) lowers the VMC energy.
    Increasing the size of the counting Jastrow basis by instead placing ellipsoidal counting functions (eFD6) around each atom did not consistently improve the variational energy for reasons discussed in section \ref{sec-normalized_counting_functions}.
    }
	\label{fig-c2h4_ng}
\end{figure}

\begin{figure}
    \centering
	\includegraphics[width=3.33in]{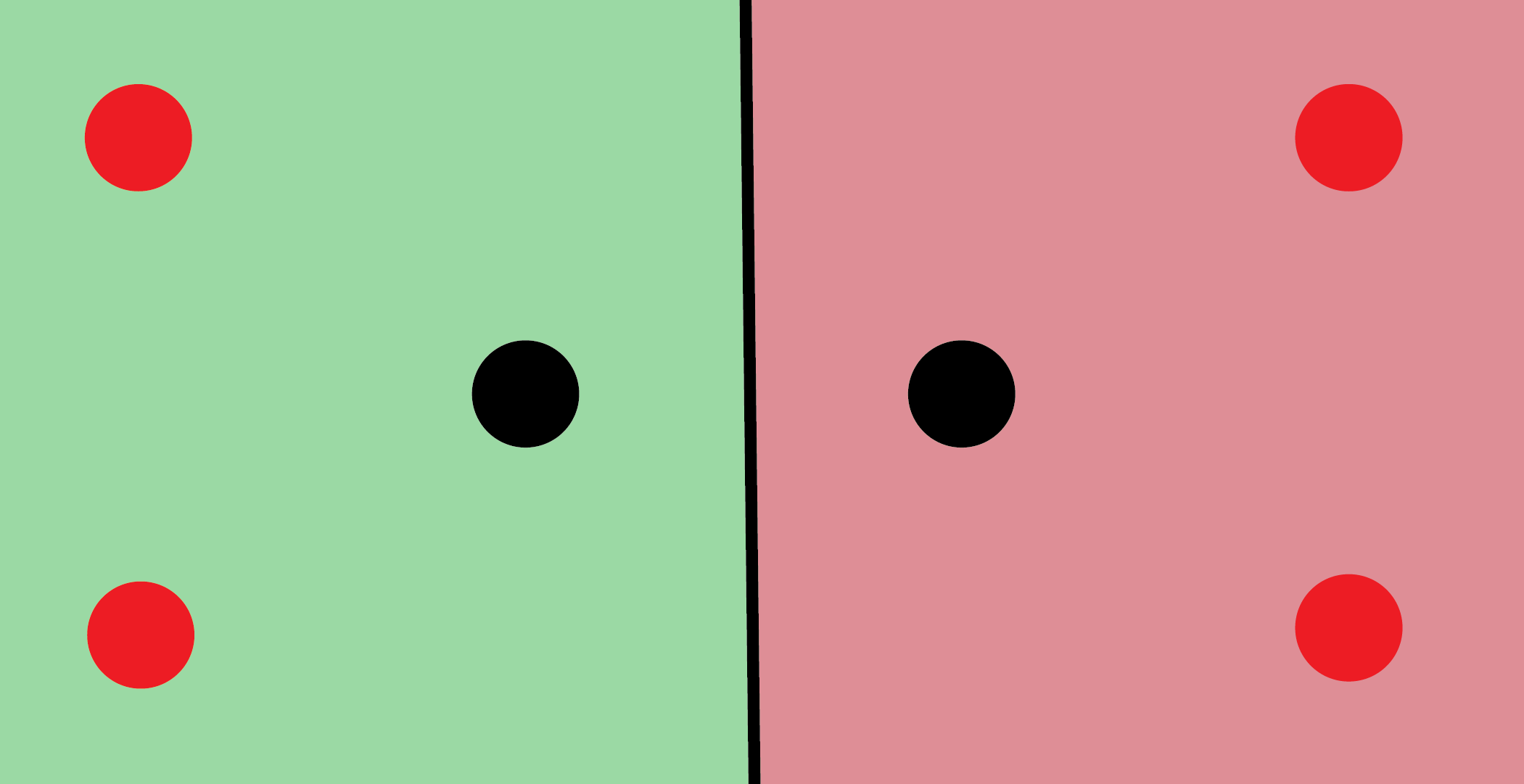}
	\caption{Illustration of the counting regions used in the NG2-CJS and FD2-CJS wavefunctions within the plane of the ethene molecule.
    Carbon and hydrogen atoms are shown as solid black and red circles respectively. 
    Each colored regions corresponds to the interior of a single counting region and approximate region boundaries are indicated by the solid black lines.
    Regions are generated either by explicitly constructing a planar counting function (equation \ref{eq-linear_cf}) with a switching surface that bisects the central double bond (FD2) or by a two-region Voronoi tessellation using the carbon centers as Voronoi points (NG2). }
	\label{fig-c2h4_2ng_basis}
\end{figure}

\begin{figure}
    \centering
	\includegraphics[width=3.33in]{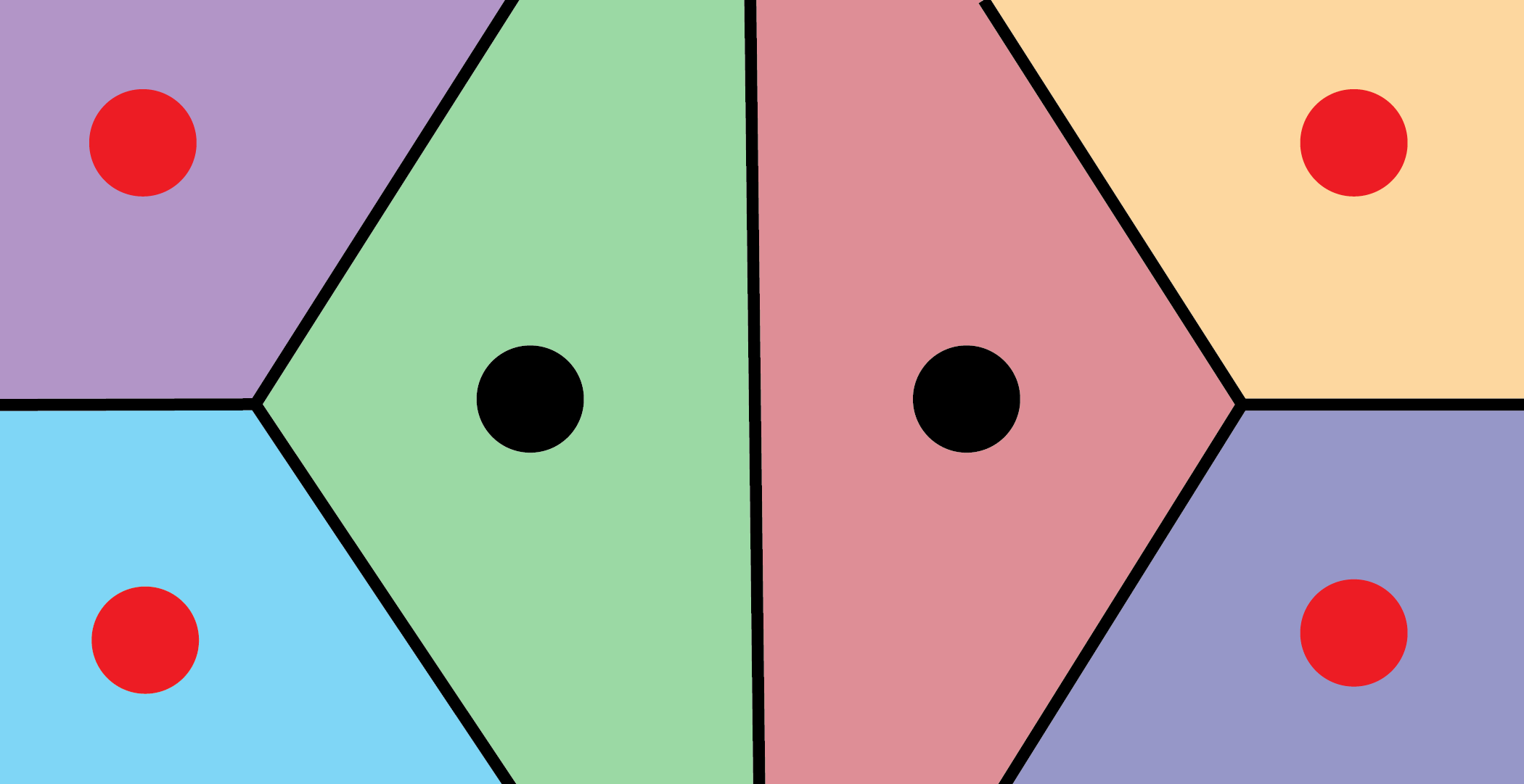}
    \caption{Illustration of the counting regions used in the NG6-CJS wavefunctions within the plane of the ethene molecule.
    Regions are built using the scheme outlined in section \ref{sec-classical_voronoi} which generates a Voronoi tessellation with atomic coordinates as the Voronoi points.}
	\label{fig-c2h4_6ng_basis}
\end{figure}

\begin{figure}
    \centering
	\includegraphics[width=3.33in]{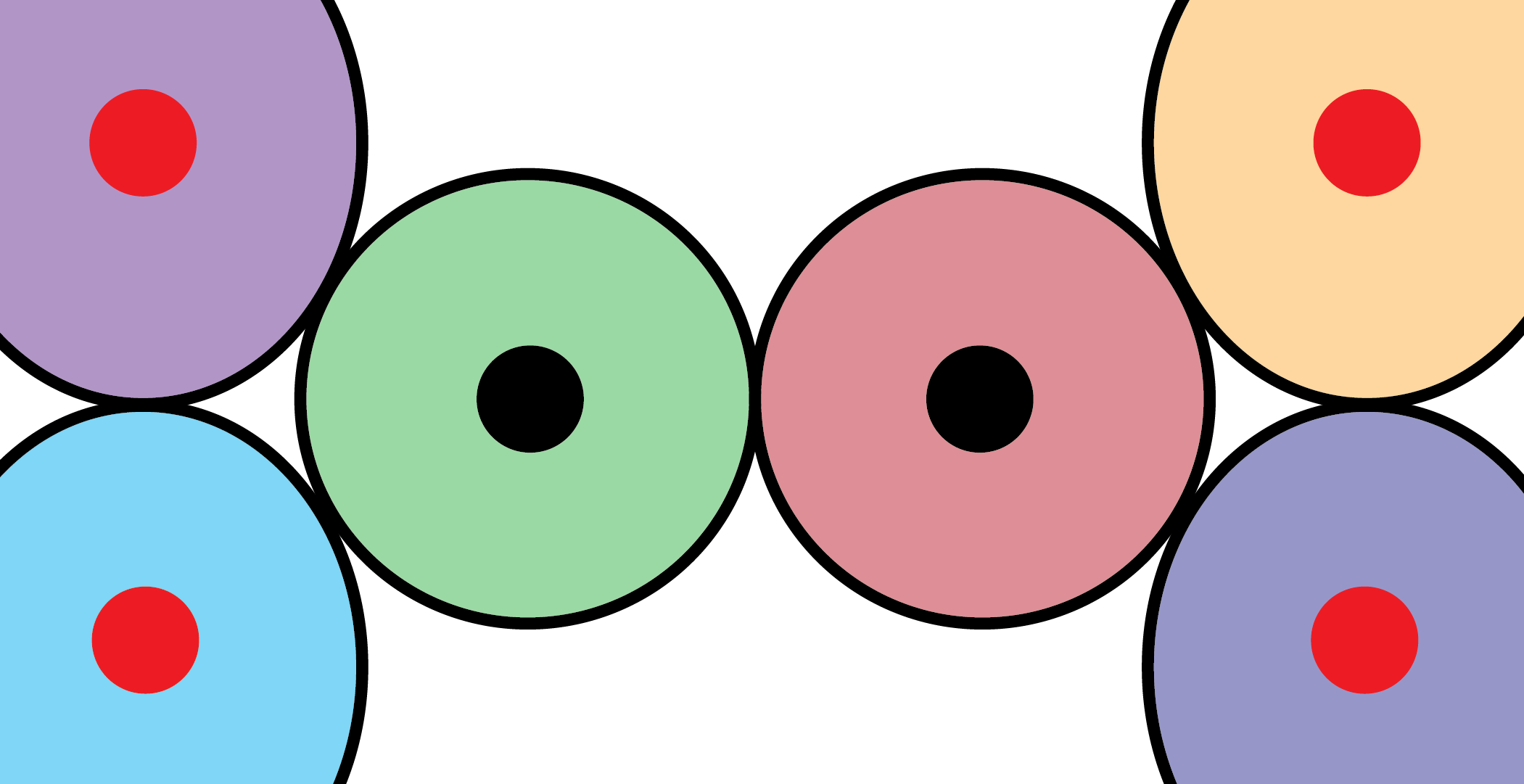}
    \caption{Illustration of the counting regions used in the eFD6-CJS wavefunctions within the plane of the ethene molecule.
    Regions are chosen to be packed ellipsoids using the Fermi-Dirac style ellipsoidal counting functions in equation \ref{eq-quad_cf}. As discussed in section \ref{sec-normalized_counting_functions}, the sigmoidal switches that occur between counting region interiors and the void regions disrupt number projections by introducing a kinetic energy cost at the boundary.}
	\label{fig-c2h4_e6fd_basis}
\end{figure}

We had previously shown\cite{vdg_ncjf} that a pair of bipartite planar counting functions recovered the correct energy ordering associated with nodal surfaces between orbitally-localized and orbitally-delocalized single-determinantal wavefunctions of ethene dissociating symmetrically along the carbon-carbon bond.
The nearly-indistinguishable dissociation curves obtained from FD-CJS and NG2-CJS ansatzes in figure \ref{fig-c2h4_ng} demonstrates the functional equivalence of an explicit bipartite basis given in equation \ref{eq-quad_cf} and counting functions generated from appropriately parameterized carbon-centered normalized gaussians given in equation \ref{eq-normgauss}. 
The counting functions used in each of these wavefunctions are schematically depicted in figures \ref{fig-c2h4_2ng_basis}, \ref{fig-c2h4_6ng_basis}, and \ref{fig-c2h4_e6fd_basis}.

Using six atom-centered Voronoi counting regions built using the scheme in section \ref{sec-classical_voronoi}, and depicted in figure \ref{fig-c2h4_6ng_basis}, the NG6-CJS wavefunction achieves a lower variational energy than its two-region counterpart. 
In the NG6-CJS wavefunction, a contraction of the three counting functions that lay on the same methylene fragment behaves much like the counting functions in the NG2-CJS wavefunction, as the counting function gradient and curvature in the interior of the fragment cancels between neighboring regions due to the property of pointwise additivity described in section \ref{sec-normalized_counting_functions}.
Thus, the NG2-CJS counting function basis is (nearly exactly) contained in the space spanned by the NG6-CJS counting function basis, and the NG6-CJS wavefunction strictly 
improves upon the variational freedom present in the NG2-CJS wavefunction.
As a result, we expect that the VMC energy of the NG6-CJS wavefunction to be bounded above by the VMC energy of the NG2-CJS wavefunction, which we do observe in figure \ref{fig-c2h4_ng}.
We can directly attribute this variational improvement to the participation of the hydrogen-centered counting functions in the NG6-CJS wavefunction, showing that these normalized gaussian counting functions can easily accommodate correlations between atoms in a single molecular fragment, a feat that was not achievable using the previous basis scheme due to the pathologies described in section \ref{sec-review}.
The same logic applies as we produce increasingly granular subdivisions of counting regions that strictly expand the span of the linear counting function space, and doing so systematically approaches the basis set limit for these Jastrow factors in a variational way.

\subsection{Random Planar H\textsubscript{4}}
\label{sec-random_h4}
The bipartite counting function basis in earlier calculations\cite{vdg_ncjf} had been set up in an \emph{ad hoc} way, and consisted of two counting functions whose collective planar switching surface was carefully placed to bisect a dissociating chemical bond.
As shown in section \ref{sec-classical_voronoi}, an atom-centered Voronoi tessellation can be easily generated using a normalized counting function basis described in section \ref{sec-normalized_counting_functions} by placing the anchor gaussian means at atomic coordinates.
To investigate the effectiveness of this automatic generation scheme under less controlled conditions, we look at the fractional correlation energy recovered in various wavefunctions for $94$ random planar geometries of H\textsubscript{4}, with each atomic coordinate chosen randomly and uniformly within a $5\mbox{\AA} \times 5\mbox{\AA}$ square. 
Geometries in which any two atoms were closer than $0.1\mbox{\AA}$ were discarded from a total random sample of $100$.

\begin{figure}
    \centering
	\includegraphics[width=3.33in]{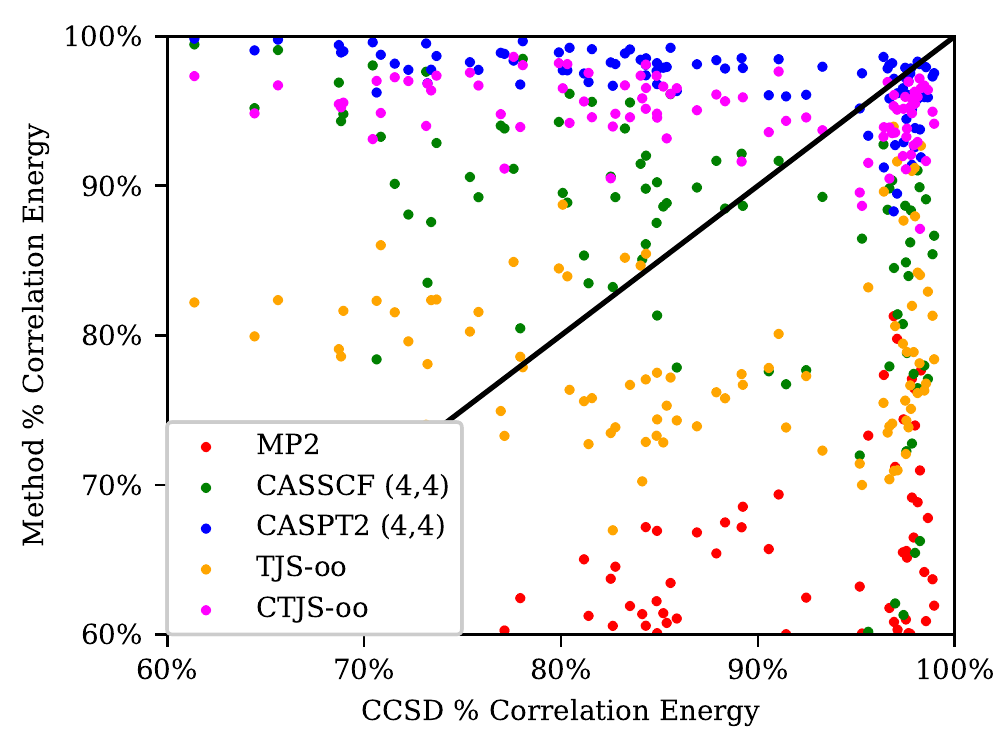}
	\caption{Fractional correlation energies recovered by various wavefunctions plotted against the fractional correlation energy recovered by a CCSD reference for 94 random arrangements of $H_4$ in in a $5\mbox{\AA} \times 5\mbox{\AA}$ plane.
    Correlation energy values for the TJS-oo and CTJS-oo wavefunctions were calculated using Variational Monte Carlo (VMC), where the absolute stochastic error is less than 1 mE\textsubscript{h} and the average error of fractional correlation energy is 0.2\%, less than the plotted symbol size.
    Calculations were performed in the cc-pVQZ basis and all wavefunction parameters were optimized in variational calculations.
    CASSCF calculated were performed using a minimal (4e,4o) active space and subsequent CASPT2 and MRCI+Q calculations considered all single and double excitations from this space.
    The counting regions used in CTJS-oo wavefunctions were generated according to section \ref{sec-classical_voronoi} in an atom-centered Voronoi arrangement, and 
    an example counting region setup is given in figure \ref{fig-h4_diagram}.
     Benchmark correlation energies were calculated by performing a three-point (cc-pVDZ, cc-pVTZ, cc-pVQZ) basis set extrapolation on MRCI+Q energies\cite{dunning_basis, varandas_basisextrap}.}
	\label{fig-h4_correlation}
\end{figure}

\begin{figure}
    \centering
	\includegraphics[width=3.33in]{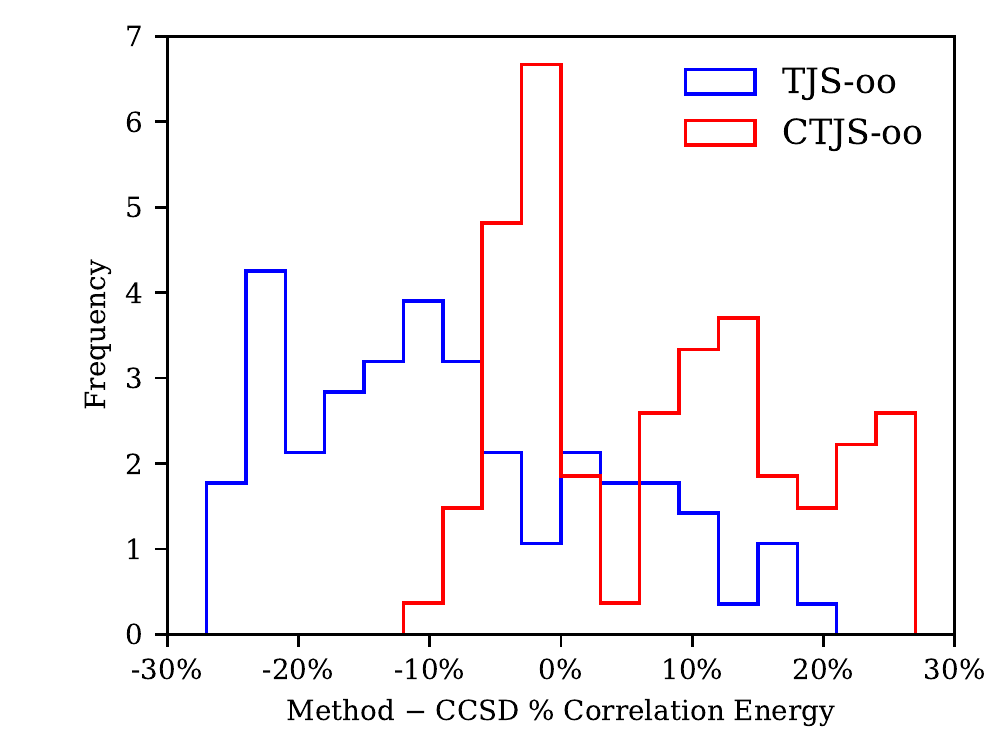}
	\caption{Frequency of fractional correlation energies recovered by the TJS-oo and CTJS-oo wavefunctions relative to a CCSD reference, using the data set plotted in figure \ref{fig-h4_correlation}. Fractional correlation energies for both wavefunctions were calculated using VMC, and the average fractional correlation energy error is 0.2\%.}
	\label{fig-h4_correlation_2}
\end{figure}

Fractional correlation energies recovered by various methods are compared to those obtained from CCSD in figure \ref{fig-h4_correlation}, and indicate that these counting Jastrow factors recover a significant fraction of the total correlation energy in geometries where the other single-reference methods (TJS-oo and CCSD) struggle to do so.
Addition of the counting Jastrow notably improves the correlation energy distribution relative to the standard TJS-oo wavefunctions across random geometries, shown in figure \ref{fig-h4_correlation_2}, and consistently recovers a high fraction of correlation energy on par with standard deterministic multireference methods in this simple system, as shown in table \ref{tab-ecorr_stats}.

\begin{table}
\centering
\caption{Mean and variance of the fractional correlation energies of 94 random H\textsubscript{4} geometries using the data set in \ref{fig-h4_correlation}.  }
\begin{tabular}{c|S[table-format=2.1]S[table-format=2.1]}
Wavefunction &  {\% $\text{E}_{\text{corr}}$ Average }  & {$\% \text{E}_{\text{corr}}$ Std. Dev.} \\
\hline
MP2 & 64.3\% & 6.1\% \\
TJS-oo  & 78.6\% & 5.7\%  \\
CASSCF(4e,4o) & 84.0\% & 12.5\% \\
CCSD & 87.1\%  & 10.4\%  \\
CTJS-oo & 95.0\% & 2.2\% \\
CASPT2 & 97.0\% & 2.3\% \\
\end{tabular}
\label{tab-ecorr_stats}
\end{table}

\begin{figure}[b]
    \centering
	\includegraphics[width=2.50in]{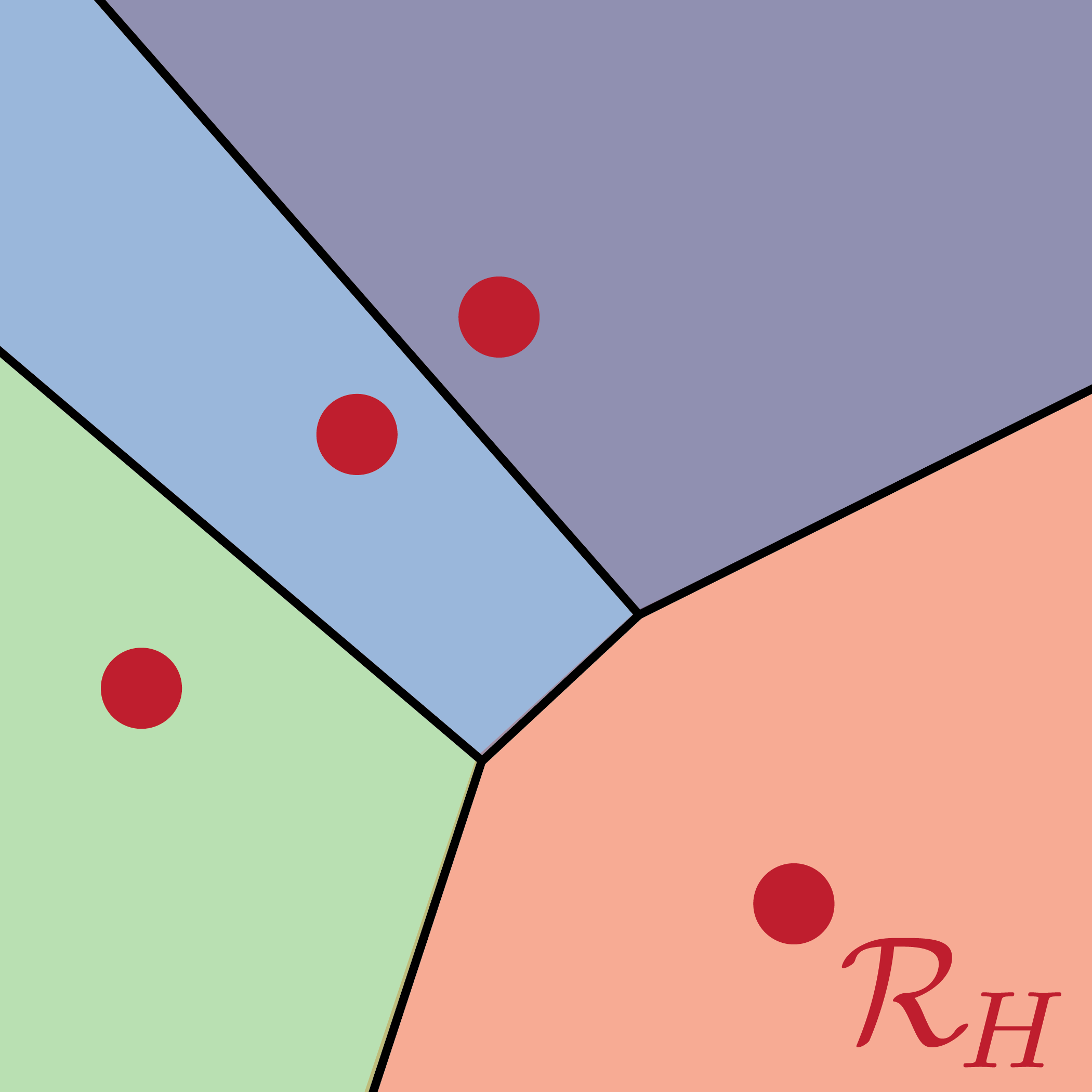}
	\caption{A representative random arrangement of H\textsubscript{4} atoms (indicated by solid red circles) and counting regions (indicated by colored areas) in a $5\mbox{\AA}\times 5\mbox{\AA}$ plane. 
    The population densities of the orange counting function $\mathcal{R}_H$ in the lower-right of the figure are calculated using equation \ref{eq-population_spectrum} for several variationally optimized wavefunctions  in figure \ref{fig-h4_spectra}.}
	\label{fig-h4_diagram}
\end{figure}

We can directly measure the projective effect of these Jastrows by plotting the density of region populations within the many-body wavefunction:
\begin{equation}\label{eq-population_spectrum} \rho_I(N) = \int |\Psi(\{\vec{r}_i\})|^2 \ \delta\left(N - \sum\limits_i C_I(\vec{r}_i)\right) d\{\vec{r}_i\}\end{equation}
The function $\rho_I(N)$ is the density of electronic configurations in $|\Psi|^2$ for which the counting region $\mathcal{R}_I$ contains $N$ electrons as determined by the counting function $C_I$.

\begin{figure}[b]
    \centering
	\includegraphics[width=3.33in]{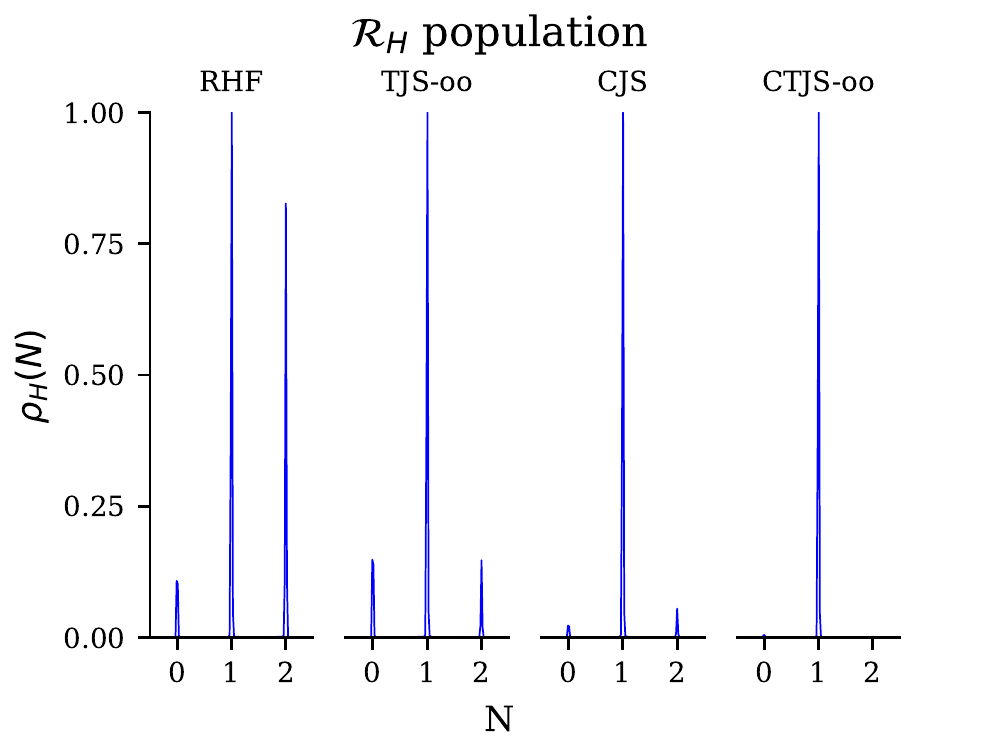}
	\caption{Sampled population density of the counting region $\mathcal{R}_H$ from figure \ref{fig-h4_diagram} for various real-space wavefunctions using equation \ref{eq-smooth_spectrum} with $\sigma = 0.03$.
    Narrow peak widths indicate both that there is little overlap between the reference wavefunction and the counting function switching surface and that this counting function acts much like a Fock-space number operator.
    The single-determinant reference contains high-energy ionic terms that are only partially removed by the one- and two-body Jastrow factors present in a TJS-oo wavefunction but are almost completely removed by the
    counting Jastrow factor in the CJS and CTJS-oo wavefunctions.
    }
	\label{fig-h4_spectra}
\end{figure}

The population density of the counting function $\mathcal{R}_H$ in the atomic geometry shown in figure \ref{fig-h4_diagram} is plotted in figure \ref{fig-h4_spectra}.
The value of this density function is approximated in a continuous way by stochastically sampling the wavefunction probability distribution $|\Psi|^2$, approximating the integral in equation \ref{eq-population_spectrum} as a sum over these samples, and smoothing the resulting distribution using fixed-width gaussians:
\begin{align}
&\rho_I(N) \approx \notag \\
&\frac{1}{\nu\sqrt{2\pi\sigma^2}} \sum\limits_{\{\vec{r}_i\} \sim |\Psi|^2}^{\nu}
\exp\left[ -\left(N - \sum\limits_i C_I(\vec{r}_i)\right)^2/(2\sigma^2) \right]
\label{eq-smooth_spectrum}
\end{align}
The sampling procdure can be easily integrated into existing optimization techniques which rely on wavefunction derivatives because the counting function values directly correspond to the derivatives of linear counting Jastrow coefficients:
\begin{equation} \frac{\partial J}{\partial G_I} = \sum\limits_i C_I(\vec{r}_i) \end{equation}

Comparing the population density peak-heights between different wavefunctions reveals that the counting Jastrow factor suppresses ionic configurations in ways that one- and two-body Jastrow factors cannot. 
Ionic configurations corresponding to the peaks at $N=0$ and $N=2$ are only partially removed in the TJS wavefunction but more fully projected out in CJS wavefunctions using a simple atom-centered Voronoi basis.
Based on the counting Jastrows' effectiveness when applied to these randomly generated systems, we conclude that these normalized counting function basis has potential as a black-box number projection factor in more general chemical systems, and are primed to perform number projections when variationally favorable.

\subsection{Calcium Oxide}
\label{sec-calcium_oxide}

Our calculations have thus far focused on simple examples where the number projecting action of the counting Jastrow is applied to remove ionic terms from a symmetry-restricted single-reference wavefunction.
A minimal, atom-centered Voronoi basis (section \ref{sec-classical_voronoi}) of counting functions is sufficient to correctly dissociate the hydrogen molecule, recover the correct state nodal surface for ethene at dissociation\cite{vdg_ncjf}, and account for strong correlation in randomized planar H\textsubscript{4} geometries.
However, in non-symmetric dissociation processes, fragment-based number projection accomplishes little, as molecular orbitals tend to already localize in SCF algorithms to avoid producing high-energy interfragment ionic terms.
Table \ref{tab-cao_vmc} demonstrates this effect prominently in the dissociation of molecular calcium oxide, where the 2-CJS wavefunction containing a counting Jastrow built from a pair of atom-centered counting functions fails to meaningfully improve the wavefunction's energy relative to the Hartree-Fock reference.

Section \ref{sec-normalized_counting_functions} shows how normalized counting functions can tile space in systematic, automatic, and flexible ways while avoiding kinetic energy problems
that result from careless placement of switching surfaces.
Normalized counting functions are built from a set of anchor gaussians whose parameters can be chosen to divide space into classical (section \ref{sec-classical_voronoi}) or spherical (section \ref{sec-spherical_voronoi}) Voronoi grids, and are not limited to the simple bipartite partitions previously used.
As described in section \ref{sec-region_composition}, smaller grids may be embedded within  a single larger grid so that the resulting patterns of counting regions preserves the structure of both grids simultaneously.
Subdividing atomic Voronoi regions into spherical sections using this composition scheme gives counting Jastrows the ability to describe population correlations within and between atomic centers simultaneously.

Molecular calcium oxide undergoes a complex dissociation, and exhibits both a singlet-triplet crossing very near equilibrium geometries and a valence calcium d-orbital participating in the chemical double bond.
In order to address the complex electronic correlations during the dissociation process, we embed a 16-region spherical partition within each of two atom-centered regions.
These spherical partitions divide the space around each of the calcium and oxygen atoms into Cartesian octants and further into two radial shells, as depicted in figure \ref{fig-cao_cjf_3d}.
The composite counting regions very nearly meet the conditions discussed in appendix \ref{app-region_composition} as the counting functions each 16-region subdivision can be recombined to approximately form each of the two original atom-centered counting functions.
Much like atomic orbital exponents\cite{gaussian_opt}, the anchor gaussian exponent parameters are highly nonlinear -- a small change in these parameters potentially changes multiple switching surfaces simultaneously -- and we saw little variational benefit when optimizing them alongside other wavefunction parameters in these more complex partitions.
As a result, these anchor gaussian parameters are held fixed in order to reduce the burden of parameter optimization and to simplify the population density analysis performed afterward.
The geometric parameters of this spherical partition and the gradient values across the counting functions' switching surfaces were chosen based on those that minimized CJS variational energies for atomic beryllium, calcium, oxygen, and nitrogen whose counting functions had the same radial and angular divisions.
\begin{figure}
    \centering
	\includegraphics[width=3.33in]{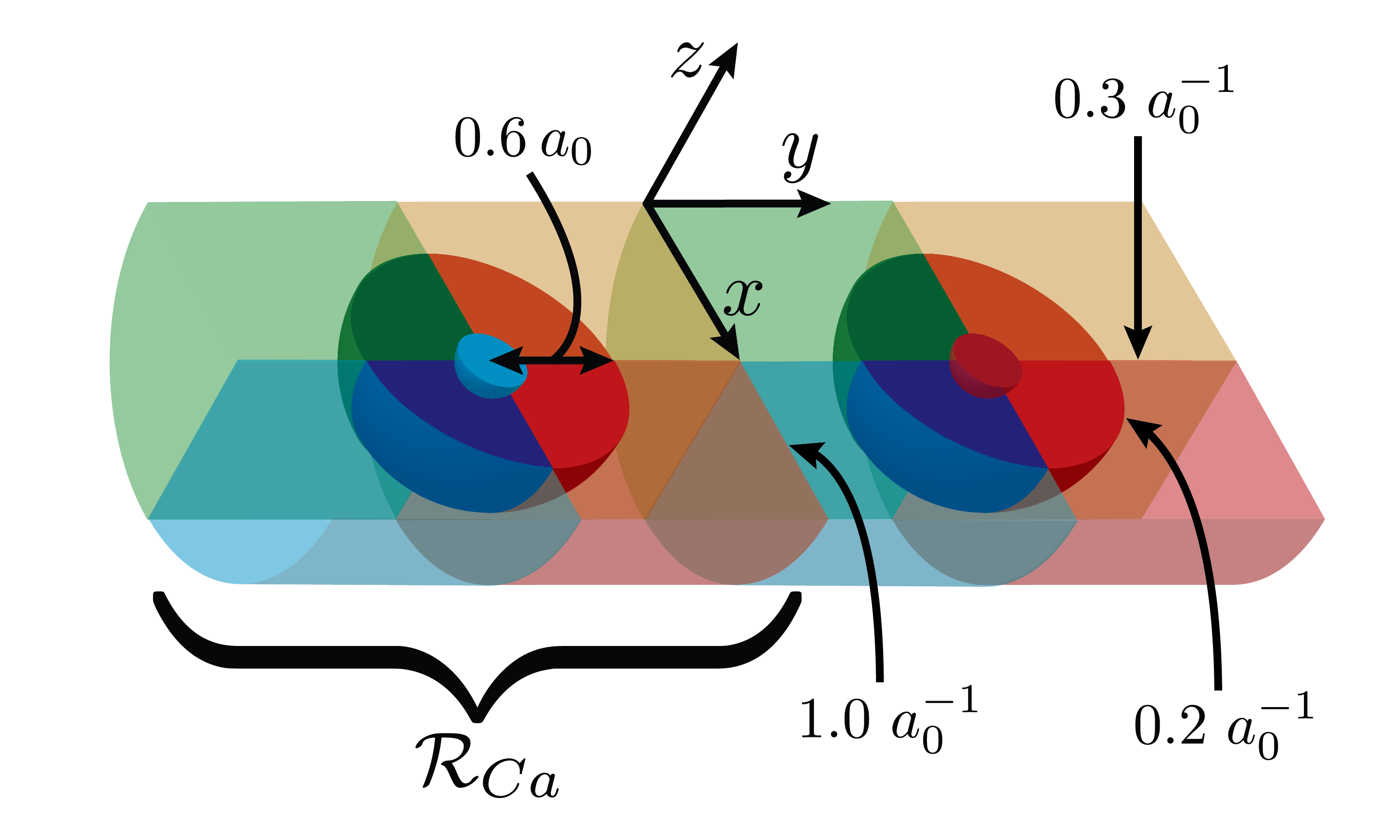} 
	\caption{Cutaway of the counting regions used in the CaO counting Jastrow factor. 
    The calcium atom is indicated by the solid blue half-sphere and the oxygen atom is indicated by the solid red half-sphere, and the remaining colored spherical sections indicate individual counting regions.
    Two atom-centered regions first divide space along a plane that bisects the bonding axis and attains a maximum slope of $1.0$ Bohr$^{-1}$ and are further subdivided into eight octants and two spherical shells to give a total of 32 counting regions. 
    Only those below the x-y plane are shown here.
    Radial boundaries of both spherical partitions occur at a radius of $0.6$ Bohr on which pair counting functions have a maximum slope of $0.2$ Bohr$^{-1}$.
    Angular switches divide each atomic region into eight octants, and attain a maximum slope of $0.3$ Bohr$^{-1}$ across their entire switching surface.
    $\mathcal{R}_{Ca}$ corresponds to the union of sixteen regions that subdivide the Calcium-centered Voronoi region, and is the focus of the population density analyses in figures \ref{fig-cao_singlet_pop}, \ref{fig-cao_singlet_pop2}, and \ref{fig-cao_triplet_pop}.
}
	\label{fig-cao_cjf_3d}
\end{figure}

In this system, we focus our attention on the accuracy of the crossing point and the energy gap between the $^{1}\Sigma$ singlet and $^{3}\Pi$ triplet states of calcium oxide at bondlengths between $2\mbox{\AA}$ and $3\mbox{\AA}$ using a single-determinant Jastrow-Slater wavefunction.
In contrast to the minimal, bipartite atom-centered Voronoi counting region partition, a counting Jastrow using the more intricate 32-region basis recovers a significant amount of correlation energy at the VMC level, and meaningfully improves both the singlet-triplet crossing point and the RMSE of the singlet-triplet gap function, as shown in table \ref{tab-cao_vmc}.
The combined effect of both cusp-correcting and counting Jastrows in the 32-CTJS wavefunction moves the crossing point approximately $0.1 \mbox{\AA}$ further toward the benchmark MRCI+Q crossing point near 2.25 \AA.
Further optimizing the orbitals in the 32-CTJS-oo wavefunctions also improves the singlet-triplet gap RMSE by a modest $15.1$ mE\textsubscript{h} relative to the more conventional TJS-oo wavefunction.

\setlength\tabcolsep{0.0pt}
\begin{table}
\caption{
    Singlet-triplet crossing points, the singlet's correlation energy E\textsubscript{corr} in mE\textsubscript{h} at 3 {\AA}, and the root-mean square error (RMSE) of the singlet-triplet gap
    in mE\textsubscript{h} calculated using VMC for a set of five Ca-O bondlengths at $0.25\mbox{\AA}$ intervals from $2\mbox{\AA}$ to $3\mbox{\AA}$ compared to an (8e,80) MRCI+Q reference.
    Number-prefixes for CJS wavefunctions correspond to the size of the counting function basis, and are either atom-centered Voronoi regions (2-CJS) or set up according to the description in figure \ref{fig-cao_cjf_3d} (32-CJS). 
    Crossing points and their uncertainties were determined by finding the roots of the singlet-triplet gap function using a curve fit given by a three-parameter exponential function $a + b\exp\left(-cx\right)$ and were found to be largely insensitive to the choice of fitting function.
    Estimates for the crossing points of the 2-CJS wavefunctions are particularly poor since, much like the bare single-determinant wavefunction, they did not exhibit a singlet-triplet crossing in the bondlength interval studied.
}
\begin{tabular}{l r @{.} l r l r @{.} l}
\hline\hline
Method &
\multicolumn{2}{ c }{ \hspace{0mm} Crossing ({\AA}) \hspace{0mm} } &
\multicolumn{2}{ c }{ \hspace{2mm} E\textsubscript{corr} \hspace{0mm} } &
\multicolumn{2}{ c }{ \hspace{2mm} Gap RMSE \hspace{0mm} } \\
\hline
2-CJS      & \hspace{2mm} 1&87(5)  & \hspace{1mm}   0&(3) & \hspace{7mm} 204&4 \\
2-CJS-oo   & \hspace{0mm} 1&94(3)  & \hspace{0mm}   6&(3) & \hspace{0mm} 188&6 \\
32-CJS     & \hspace{0mm} 2&01(4)  & \hspace{0mm} 360&(2) & \hspace{0mm}  60&4 \\
32-CJS-oo  & \hspace{0mm} 1&994(2) & \hspace{0mm} 394&(3) & \hspace{0mm}  43&4 \\
TJS        & \hspace{0mm} 2&02(2)  & \hspace{0mm} 524&(1) & \hspace{0mm}  54&5 \\
TJS-oo     & \hspace{0mm} 2&012(6) & \hspace{0mm} 546&(1) & \hspace{0mm}  43&8 \\
32-CTJS    & \hspace{0mm} 2&08(1)  & \hspace{0mm} 543&(1) & \hspace{0mm}  49&9 \\
32-CTJS-oo & \hspace{0mm} 2&081(4) & \hspace{0mm} 581&(1) & \hspace{0mm}  28&7 \\
CCSD(T)    & \hspace{0mm} 2&32(4)  & \hspace{0mm} 697&    & \hspace{0mm}  16&2\\
MRCI+Q     & \hspace{0mm} 2&250(3) & \hspace{0mm} 675&    & \multicolumn{2}{c}{\hspace{3mm}N/A} \\
\hline\hline
\end{tabular}
\label{tab-cao_vmc}
\end{table}
\setlength\tabcolsep{6.0pt}

\begin{figure}
    \centering
	\includegraphics[width=3.33in]{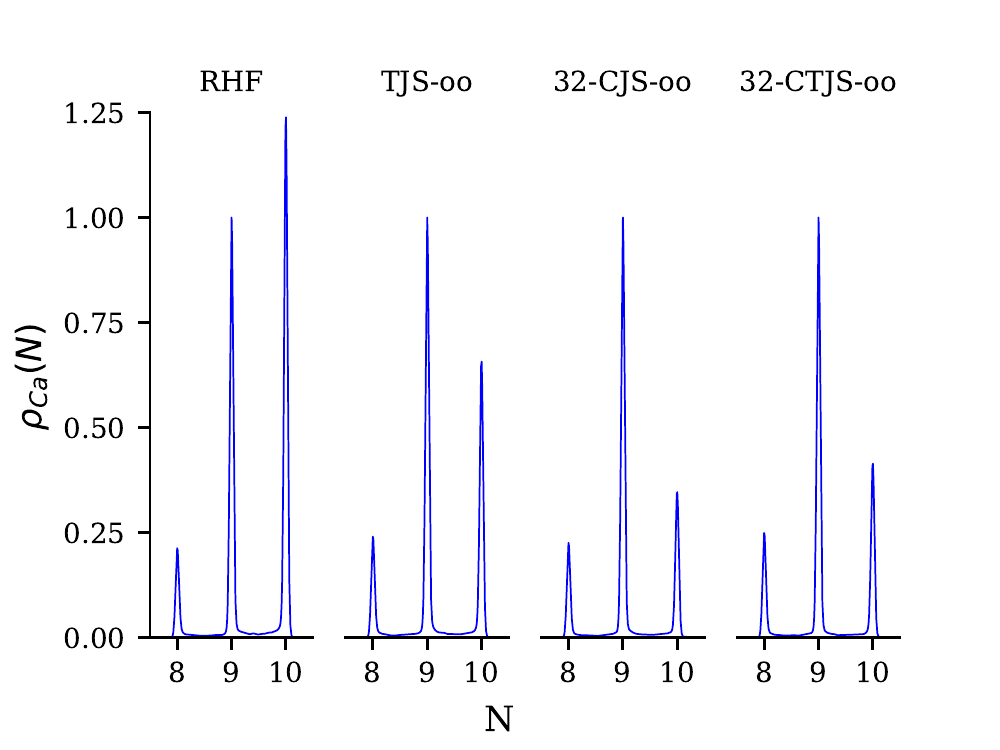}
	\caption{
        Population density distribution of the aggregate 16 regions surrounding the calcium atom indicated by $\mathcal{R}_{Ca}$ in figure \ref{fig-cao_cjf_3d} calculated using equation \ref{eq-smooth_spectrum} with $\sigma = 0.03$ for the indicated singlet wavefunctions at $3\ \mbox{\AA}$.
        Distributions are rescaled to unit peak height at $N=9$ and integrated peak areas of the normalized distribution is given in table \ref{tab-cao_int_spectrum}. }
	\label{fig-cao_singlet_pop}
\end{figure}

\begin{figure}
    \centering
	\includegraphics[width=3.33in]{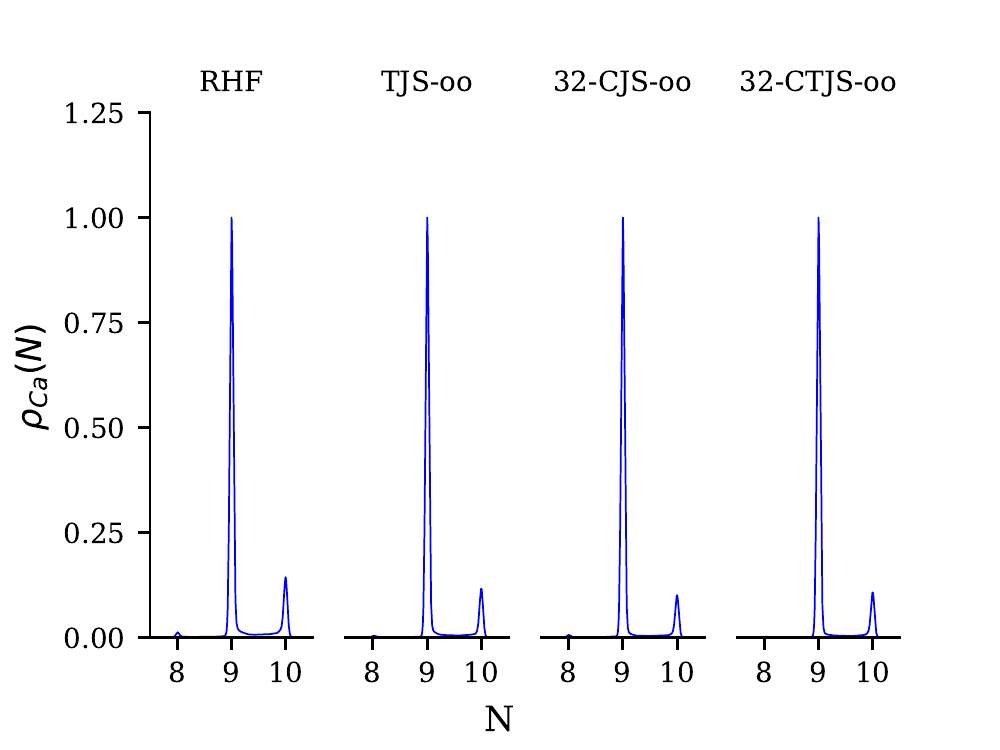}
	\caption{
        Population density distribution of the aggregate 16 regions surrounding the calcium atom indicated by $\mathcal{R}_{Ca}$ in figure \ref{fig-cao_cjf_3d} calculated using equation \ref{eq-smooth_spectrum} with $\sigma = 0.03$ for the indicated $^{3}\Pi$ triplet  wavefunctions at $3 \mbox{\AA}$.
        Distributions are rescaled to unit peak height at $N=9$ and integrated peak areas of the normalized distribution is given in table \ref{tab-cao_int_spectrum}.}
	\label{fig-cao_triplet_pop}
\end{figure}

We again look at population density functions defined in equation \ref{eq-population_spectrum} to more precisely judge the number projecting effect of the counting Jastrow factor.
Population densities of the aggregate calcium-centered counting regions, indicated by $\mathcal{R}_{Ca}$ in figure \ref{fig-cao_cjf_3d}, are given for singlet and triplet wavefunctions in figures \ref{fig-cao_singlet_pop} and \ref{fig-cao_triplet_pop} respectively.
Integrated peak areas of these population density plots -- which signify the total fraction of the wavefunction that attains the indicated electronic population in the calcium-centered region -- are given in table \ref{tab-cao_int_spectrum}.
These density functions clearly show that the counting Jastrow factors effectively redistribute electrons between atoms when applied to singlet reference configurations and act most prominently when orbitals are coupled to the counting Jastrow factor, independent of the presence of cusp-correcting Jastrow factors.
Since the two-region partition in the 2-CJS wavefunctions showed negligible effect on the wavefunction, we conclude that these are taking full advantage of this fine 32-region division to introduce intraatomic correlations that allow the Jastrow to reweigh configurations based on atomic populations.

\begin{figure}
    \centering
	\includegraphics[width=3.33in]{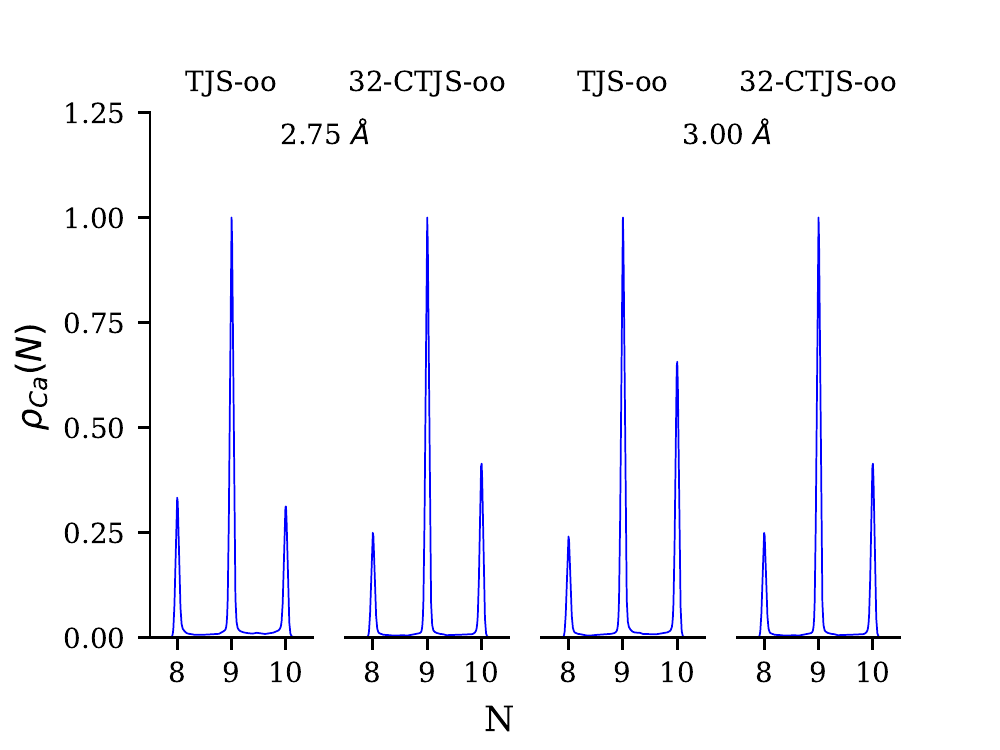}
	\caption{
        Population density distribution of the aggregate 16 regions surrounding the calcium atom indicated by $\mathcal{R}_{Ca}$ in figure \ref{fig-cao_cjf_3d} calculated using equation \ref{eq-smooth_spectrum} with $\sigma = 0.03$ for the TJS-oo and 32-CTJS-oo singlet wavefunctions at $2.75\ \mbox{\AA}$ and $3\ \mbox{\AA}$.
        Distributions are rescaled to unit peak height at $N=9$ and integrated peak areas of the normalized distribution is given in table \ref{tab-cao_int_spectrum}. 
}
	\label{fig-cao_singlet_pop2}
\end{figure}

\setlength\tabcolsep{2.0pt}
\begin{table}
\centering
\caption{Peak areas for the population density peaks of the indicated wavefunction, state symmetry, and bond distance R,
         some of which are shown in figures \ref{fig-cao_singlet_pop} and \ref{fig-cao_singlet_pop2},
         integrated over the domain $[N-0.5, N+0.5]$. 
         These values roughly correspond to the fraction of the wavefunction in which the indicated number
         of electrons populate the aggregate calcium-centered region $\mathcal{R}_{Ca}$ shown in figure \ref{fig-cao_cjf_3d}.
        }
\begin{tabular}{l l r @{.} l r @{.} l r @{.} l r @{.} l}
\hline\hline
Method &
State &
\multicolumn{2}{ c }{ \hspace{0mm} R ({\AA}) \hspace{0mm} } &
\multicolumn{2}{ c }{ \hspace{0mm} $N$=\hspace{0.4mm}8  \hspace{0mm} } &
\multicolumn{2}{ c }{ \hspace{0mm} $N$=\hspace{0.4mm}9  \hspace{0mm} } &
\multicolumn{2}{ c }{ \hspace{0mm} $N$=\hspace{0.4mm}10 \hspace{0mm} } \\
\hline
TJS-oo     & \hspace{0mm} $^{1}\Sigma$ & \hspace{0mm} 2&75 & \hspace{0mm} 0&202 & \hspace{0mm} 0&589 & \hspace{0mm} 0&204 \\
32-CJS-oo  & \hspace{0mm} $^{1}\Sigma$ & \hspace{0mm} 2&75 & \hspace{0mm} 0&199 & \hspace{0mm} 0&594 & \hspace{0mm} 0&203 \\
32-CTJS-oo & \hspace{0mm} $^{1}\Sigma$ & \hspace{0mm} 2&75 & \hspace{0mm} 0&138 & \hspace{0mm} 0&624 & \hspace{0mm} 0&222 \\
RHF        & \hspace{0mm} $^{1}\Sigma$ & \hspace{0mm} 3&00 & \hspace{0mm} 0&095 & \hspace{0mm} 0&410 & \hspace{0mm} 0&490 \\
32-CJS     & \hspace{0mm} $^{1}\Sigma$ & \hspace{0mm} 3&00 & \hspace{0mm} 0&075 & \hspace{0mm} 0&526 & \hspace{0mm} 0&396 \\
TJS-oo     & \hspace{0mm} $^{1}\Sigma$ & \hspace{0mm} 3&00 & \hspace{0mm} 0&133 & \hspace{0mm} 0&524 & \hspace{0mm} 0&336 \\
32-CJS-oo  & \hspace{0mm} $^{1}\Sigma$ & \hspace{0mm} 3&00 & \hspace{0mm} 0&147 & \hspace{0mm} 0&618 & \hspace{0mm} 0&229 \\
32-CTJS-oo & \hspace{0mm} $^{1}\Sigma$ & \hspace{0mm} 3&00 & \hspace{0mm} 0&155 & \hspace{0mm} 0&589 & \hspace{0mm} 0&251 \\
RHF        & \hspace{0mm} $^{3}\Pi$    & \hspace{0mm} 3&00 & \hspace{0mm} 0&017 & \hspace{0mm} 0&816 & \hspace{0mm} 0&161 \\
32-CJS     & \hspace{0mm} $^{3}\Pi$    & \hspace{0mm} 3&00 & \hspace{0mm} 0&023 & \hspace{0mm} 0&816 & \hspace{0mm} 0&155 \\
32-TJS-oo  & \hspace{0mm} $^{3}\Pi$    & \hspace{0mm} 3&00 & \hspace{0mm} 0&007 & \hspace{0mm} 0&861 & \hspace{0mm} 0&129 \\
32-CJS-oo  & \hspace{0mm} $^{3}\Pi$    & \hspace{0mm} 3&00 & \hspace{0mm} 0&008 & \hspace{0mm} 0&869 & \hspace{0mm} 0&115 \\
32-CTJS-oo & \hspace{0mm} $^{3}\Pi$    & \hspace{0mm} 3&00 & \hspace{0mm} 0&004 & \hspace{0mm} 0&875 & \hspace{0mm} 0&120 \\
\hline\hline
\end{tabular}
\label{tab-cao_int_spectrum}
\end{table}
\setlength\tabcolsep{6.0pt}

In contrast to VMC energies that are straightforwardly affected by the amplitudes of the trial wavefunction, Diffusion Monte Carlo (DMC) energies are related to the trial wavefunction in a more indirect way.
In order to ensure that the sampled distribution remains appropriately antisymmetric while avoiding the exponential sign problem\cite{chandrasekharan_signproblem}, DMC commonly employs the fixed-node approximation, where the fermionic nodes of a trial wavefunction restrict the Monte Carlo sampling procedure\cite{qmc_review}.
In addition, due to the smaller timestep required to resolve the higher energy scale of core electrons, nonlocal pseudopotentials replace their explicit simulation, much like the frozen core approximation that reduce the number of excitations considered in post-Hartree Fock methods.
In DMC, these nonlocal components are evaluated through the locality approximation\cite{nlpp_locality_ceperley} or its variational analogue\cite{casula_tmoves,nlpp_tmoves_filippi} (used here), and their accuracy directly relies on the quality of trial wavefunction amplitudes relative to the projected ground state.
However, the relative magnitude of these approximations remains opaque due to the complexity of both the many-body nodal surface
and nonlocal pseudopotential terms\cite{nlppe_benchmark}, and it remains difficult to differentiate or address these sources of error directly.

\begin{figure}[t]
    \centering
	\includegraphics[width=3.33in]{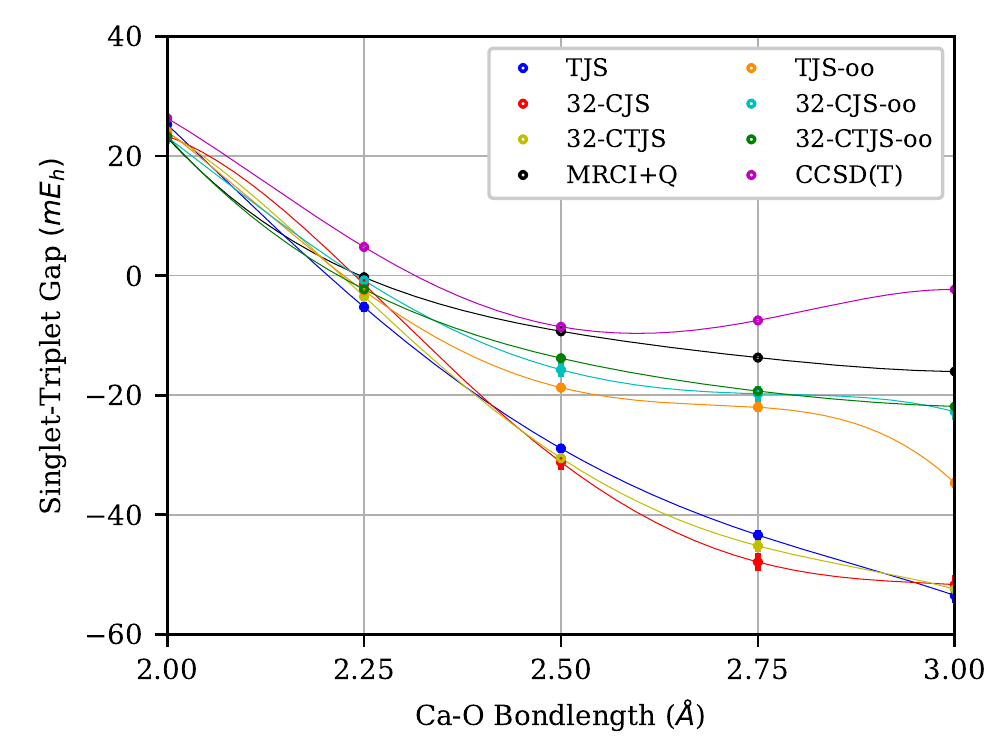}
	\caption{DMC singlet-triplet energy gap value for various wavefunctions as a function of Ca-O bondlength between $2\mbox{\AA}$ and $3\mbox{\AA}$. Lines are a guide to the eye. The DMC timestep is 0.02 $\hbar$ E\textsubscript{h}$^{-1}$ and absolute stochastic errors are less 2 mE\textsubscript{h} and are smaller than the plotted symbol size except for CJS and CJS-oo wavefunctions. }
	\label{fig-cao_dmc_gap}
\end{figure}

The behavior of the singlet-triplet DMC energy gap, shown in figure \ref{fig-cao_dmc_gap} with RMSE given in table \ref{tab-cao_dmc_gap}, roughly separates the real-space wavefunctions into two clusters: those without optimized orbitals (TJS, 32-CJS, 32-CTJS) and those with optimized orbitals (TJS-oo, 32-CJS-oo, 32-CTJS-oo)
Despite their disparate VMC energies (table \ref{tab-cao_vmc}), those in the former cluster exhibit very similar DMC energy gaps between $2.0\ \mbox{\AA}$ and $3.0\ \mbox{\AA}$.
Since these wavefunctions all share the same fermionic nodal surface, this strongly suggests that that cusp-correcting and counting Jastrow factors largely affect nonlocal pseudopotential evaluation uniformly in both singlet and triplet wavefunctions.
Note that the difference in singlet correlation energy between the 2-CJS and 32-CTJS wavefunctions gives us an idea of the magnitude of the nonlocal pseudopotential error present, as the two-region counting Jastrow does very little to change the underlying Hartree-Fock reference.
At around 60 mH, this error is surprisingly large, and yet the TJS and 32-CJS wavefunctions 
correct it to a similar degree (56 mH and 50 mH, respectively) and in similar ways (missing only 3 mH and 9 mH from the composite 32-CTJS result, respectively) despite their very different functional forms.
By contrast, those in the latter cluster exhibit a marked improvement in the RMSE of the singlet-triplet gap, and as orbital rotations provide the only avenue for changing the wavefunction's nodal surface, it is safe to say these orbital rotations -- through coupling to the Jastrow factors -- are  responsible for improvements to the nodal surface.

All of the DMC singlet-triplet gaps roughly match the MRCI+Q reference until the crossing point around $2.2\mbox{\AA}$, after which they start to deviate more strongly.
For example, while the highly accurate single-reference Fock space wavefunction CCSD(T) is accurate near equilibrium, it overestimates the crossing and starts to diverge at longer bondlengths, producing a qualitatively incorrect singlet-triplet gap at stretched geometries.
Interestingly, the inaccuracy of the DMC gap of the TJS-oo wavefunction at $3.0\ \mbox{\AA}$ is coincident with a stark change in the calcium population density, shown in figure \ref{fig-cao_singlet_pop2} and table \ref{tab-cao_int_spectrum}, a feature not shared by either the 32-CJS-oo or 32-CTJS-oo wavefunction.
In addition, the singlet-triplet gap of the 32-CJS-oo wavefunction and the 32-CTJS-oo wavefunction are nearly on top of each other despite the much higher VMC energy of the former (table \ref{tab-cao_vmc}).
These observations suggest that the counting Jastrow is particularly effective at coupling to the orbital parameters in this case, which is seemingly linked to their ability to describe long-range population correlations missing from cusp-correcting Jastrows.

\setlength\tabcolsep{0.0pt}
\begin{table}
\caption{As table \ref{tab-cao_vmc}, but for DMC instead of VMC.}
\begin{tabular}{l r @{.} l r l r @{.} l}
\hline\hline
Method &
\multicolumn{2}{ c }{ \hspace{0mm} Crossing ({\AA}) \hspace{0mm} } &
\multicolumn{2}{ c }{ \hspace{2mm} E\textsubscript{corr} \hspace{0mm} } &
\multicolumn{2}{ c }{ \hspace{2mm} Gap RMSE \hspace{0mm} } \\
\hline
2-CJS      & \hspace{2mm} 2&20(3)  & \hspace{1mm} 571&(2) & \hspace{7mm} 22&9 \\
2-CJS-oo   & \hspace{0mm} 2&220(4) & \hspace{0mm} 569&(2) & \hspace{0mm} 23&3 \\
32-CJS     & \hspace{0mm} 2&19(3)  & \hspace{0mm} 621&(1) & \hspace{0mm} 24&1 \\
32-CJS-oo  & \hspace{0mm} 2&22(5)  & \hspace{0mm} 651&(1) & \hspace{0mm}  4&9 \\
TJS        & \hspace{0mm} 2&21(1)  & \hspace{0mm} 627&(1) & \hspace{0mm} 23&2 \\
TJS-oo     & \hspace{0mm} 2&19(5)  & \hspace{0mm} 647&(1) & \hspace{0mm} 10&1 \\
32-CTJS    & \hspace{0mm} 2&19(2)  & \hspace{0mm} 630&(1) & \hspace{0mm} 23&6 \\
32-CTJS-oo & \hspace{0mm} 2&220(1) & \hspace{0mm} 664&(1) & \hspace{0mm}  4&2 \\
CCSD(T)    & \hspace{0mm} 2&32(3)  & \hspace{0mm} 697&    & \hspace{0mm} 16&2 \\
MRCI+Q     & \hspace{0mm} 2&249(3) & \hspace{0mm} 675&    & \multicolumn{2}{c}{\hspace{2mm}N/A} \\
\hline\hline
\end{tabular}
\label{tab-cao_dmc_gap}
\end{table}
\setlength\tabcolsep{6.0pt}

\section{Conclusions and Outlook}
\label{sec-conclusions}
In this paper, we have shown how to construct a one-particle basis which partitions space in a complete and natural way. 
These basis functions are referred to as counting functions, and are designed to act as the position-space equivalent of Fock-space number operators.
When used as the basis in a four-body Jastrow correlation factor these counting functions are able to correlate electronic populations between well-characterized counting regions. 
These counting functions are smooth, sigmoidal functions that are made by normalizing a set of three-dimensional gaussian functions and are complete, localizable, and cleanly additive.
We show how this normalization condition is responsible for these attractive formal properties, and how it alleviates many of the problems posed by a real-space formulation of a population-based projection factor.
Simple sigmoidal functions closely approximate local facets of these counting regions, and the boundaries of these regions may be clearly visualized as a patchwork of linear and quadratic surfaces.
We provide two parameterization schemes designed to describe correlations in orbital populations between and within atoms that arrange counting regions in patterns of atom-centered Voronoi cells or extruded spherical Voronoi partitions respectively. 
Finally, taking advantage of these basis functions' clean additivity, we show how a composition scheme can be used to subdivide these counting functions in a useful way.

In simple molecular systems, these number projections, alongside cusp-correcting Jastrows and orbital rotations, capture nodal surface details and correlation energies beyond what is currently achievable using several sophisticated single-reference methods.
For example, in random planar arrangements of H\textsubscript{4}, counting-Jastrow-augmented wavefunctions consistently outperform CCSD and TJF-oo, and on average recovers nearly as much fractional correlation energy as multi-determinant CASPT2, and can be optimized at a computational cost that scales no higher than existing real-space Jastrows.
For nonsymmetric dissociations, like molecular calcium oxide, a simple pair of atom-based Voronoi counting regions are ineffective at improving the wavefunction through the counting Jastrow.
A counting function basis that subdivides each of these atom-based cells into a coarse spherical grid meaningfully improves the 32-CJS-oo variational energies, and reduces the mean square error of the DMC singlet-triplet gap around the crossing point by a factor of two relative to the TJF-oo wavefunction and three relative to CCSD(T).
The counting Jastrows' number projecting action is explicitly verified by inspecting measures of the counting region population density, and appears to be more effective at suppressing molecular terms with high charge-separation than standard real-space jastrow factors in H\textsubscript{4} and CaO.

As a multiplicative real-space factor, counting Jastrow factors are trivially compatible with existing real-space wavefunctions, and have the potential to augment them in a unique and compact way.
Our present calculations have used only a single-determinant reference in an effort to demonstrate the power of these number projecting factors in a limited setting, and future work will explore complex references that span a larger portion of the configuration space, taking full advantage of recent advances in multi-slater expansion optimization techniques\cite{sergio_multislater,claudia_multislater} and compact functional forms like the antisymmeterized geminal power ansatz.
The counting Jastrow does not increase the compuational cost-scaling of wavefunction evaluations compared to those already within the reach of DMC in larger chemical systems (such as TJS with $\sim$1000 electrons)\cite{bigdmc_1,bigdmc_2,bigdmc_3} and can be straightforwardly applied at scale, subject to the development of optimization techniques\cite{zhao2017blocked} that can efficiently handle the growing number of variational parameters that these Jastrows and orbital rotations introduce.
In particular, the counting Jastrow's efficient representation of population-based correlations suggests application to charge-transfer excited states and complements recent work in excited state variational theory\cite{shea_excited}.
As we look toward applying these counting Jastrows in more complex systems, the flexibility and adaptability built into this normalized gaussian basis will allow us to systematically construct compact and powerful number projections and optimize them at low-order polynomial cost.

\section{Acknowledgments}
This work was supported by funding from the Early Career Research Program of the Office of Science, Office of Basic Energy Sciences, the U.S. Department of Energy, Grant No. {DE-SC0017869}.
Calculations were performed partly on the UC Berkeley Savio computing cluster and partly at the National Energy Research Scientific Computing Center,
a DOE Office of Science User Facility supported by the Office of Science of the U.S. Department of Energy under Contract No. {DE-AC02-05CH11231}. 

\appendix
\section{Linear Dependencies}
\label{app-linear_dependencies}

The Linear Method \cite{linearmethod1,linearmethod2} is a wavefunction minimization algorithm that begins by expanding the wavefunction into its first-order derivative, or tangent, space:
\begin{equation} \left| \Psi_{lin}\right> = d_0\ket{\Psi(\vec{p})} + \sum\limits_{i} d_i\ket{\frac{\partial \Psi(\vec{p})}{\partial p_i}}  \end{equation}
As wavefunctions at an energy minimum will necessarily have a vanishing gradient, we solve for the linear coefficients $d_i$ by setting:
\begin{equation}\nabla_{\vec{d}} \left( \frac{\left< \Psi_{lin} \right| \mat{H} \left| \Psi_{lin} \right>}{\left< \Psi_{lin} \right|\left. \Psi_{lin} \right>} \right) = 0 \end{equation}
which becomes a generalized eigenvalue equation whose lowest-energy eigenvector is used to update the wavefunction:
\begin{equation} \mat{H}^{lin} \vec{d} = E^{lin} \mat{S}^{lin} \vec{d}, \quad \ket{\Psi(\vec{p})} \rightarrow \ket{\Psi(\vec{p} + \vec{d})} \end{equation}
Matrix elements are given by the expectation values of tangent space components:
\begin{align}
\mat{H}^{lin}_{ij} &= \left< \frac{\partial \Psi(\vec{p})}{\partial p_i} \right| \mat{H} \left| \frac{\partial \Psi(\vec{p})}{\partial p_j} \right> \notag \\
\mat{S}^{lin}_{ij} &= \left< \frac{\partial \Psi(\vec{p})}{\partial p_i} \right| \left. \frac{\partial \Psi(\vec{p})}{\partial p_j} \right>
\end{align}
As often happens in numerical linear algebra, the solutions to this generalized eigenvalue equation become ill-conditioned as the eigenvalues of the overlap matrix $\mat{S}^{lin}$ approach zero.
The numerical properties of the overlap matrix are strongly intertwined with the basis used to construct it, and becomes singular whenever this basis is not linearly independent.
We will show exactly how redundant representations of the wavefunction in parameter space lead to linear dependencies in its tangent space basis, creating a singular overlap matrix.
Though the normalization condition is the source of many of the formal benefits enjoyed by the basis introduced in section \ref{sec-normalized_counting_functions}, it also produces exactly this type of parameter redundancy to the counting Jastrow factor.
As a result, the identification and removal of excess degrees of freedom is crucial to obtain numerically well-behaved linear method equations when optimizing these counting Jastrow factors.

In this section, $\Phi$ represents a mapping from parameter space to the wavefunction Hilbert space:
\begin{equation} \Phi: \mathbb{R}^n \rightarrow \mathbf{H}\end{equation}
First, we will assume that this mapping is continuous, and that there exists a continuous, nontrivial `equivalence path' $\vec{p}$ along which $\Phi$ maps to the same wavefunction to within a normalization factor. 
For example, for:
\begin{equation} \braket{x|\Phi(a,b)} = \exp\left(-a b x^2\right)\end{equation}
this path would be:
\begin{equation} \vec{p}(\alpha) = a b \left[ \begin{array}{c} \alpha \\ 1/\alpha \end{array} \right], \quad \alpha \in (0,\infty) \end{equation}
so that:
\begin{equation} \Phi(\vec{p}(1)) \equiv \Phi(\vec{p}(\alpha)), \text{ for all } \alpha \in (0,\infty) \end{equation}

In general, the derivative of this wavefunction mapping with respect to the path coordinate $\alpha$ of any such path is the zero vector in $\mathcal{H}$:
\begin{equation} \frac{d \Phi(\vec{p}(\alpha))}{d\alpha} = \lim\limits_{\Delta\alpha \rightarrow 0} \frac{\Phi(\vec{p}(\alpha + \Delta \alpha) - \Phi(\vec{p}(\alpha))}{\Delta \alpha} = \vec{0}\end{equation}
Expanding this derivative using the multivariate chain rule produces the linear dependency:
\begin{equation} \label{eq-lindep} \vec{0} = \frac{d \Phi(\vec{p}(\alpha))}{d \alpha} = \sum\limits_{i} \frac{\partial
 \Phi(\vec{p}(\alpha))}{\partial p_i}\frac{d p_i(\alpha)}{d \alpha} \end{equation}
As this linear combination is equal to the zero function, projecting any vector in the tangent space onto it produces a value of zero:
\begin{equation} \left< \frac{\partial \Phi(\vec{p})}{\partial p_k} \bigg| \sum\limits_i \frac{\partial \Phi(\vec{p}(\alpha))}{\partial p_i} p_i'(\alpha)\right> = 0 \end{equation}
This is equivalent to the action of the overlap matrix on the vector $\vec{p}'(\alpha)$:
\begin{equation} \mat{S}^{lin} \vec{p}'(\alpha) = \vec{0}\end{equation}
showing that the overlap matrix is singular.
Removing these linear dependencies can be done by identifying a formula for the coefficients of $\vec{p}'(\alpha)$, integrating to find the path $\vec{p}(\alpha)$, and restricting parameters to stop variations along the path-coordinate $\alpha$.
Once done, this removes the offending vectors from the linear method tangent space.

Our counting Jastrow factor is given by equation \ref{eq-ncjf_def} with a normalized counting function basis from \ref{eq-normgauss} and is written:
\begin{align}
& \Phi\left(\mat{F}, \vec{G}, \{g_I\} \right) = e^{J_{C}} \\
& J_C = \sum\limits_{ijIJ} F_{IJ} C_I(\vec{r}_i) C_J(\vec{r}_j) + \sum\limits_{iI} G_I C_I(\vec{r}_i) \\
& C_I(\vec{r}_i) = \frac{g_I(\vec{r}_i)}{\sum\limits_J g_J(\vec{r}_i)} \\
& g_I(\vec{r}_i) = \exp\left( \vec{r}^T \mat{A}_I \vec{r} - 2\vec{B}_I^T \vec{r} + K_I \right)
\end{align}

The normalization condition in the definition of $C_I$ can be used to directly construct these equivalence paths for the $\mat{F}$ and $\mat{G}$ parameters. 
We first note that these parameter derivatives are given by:
\begin{equation}\frac{\partial\Phi}{\partial G_{I}} = \sum\limits_{i} C_{I}(\vec{r}_i) \Phi, \quad\frac{\partial\Psi}{\partial F_{IJ}} = \sum\limits_{ij} C_{I}(\vec{r}_i) C_{J}(\vec{r}_j) \Phi  \end{equation}
The sum of normalized counting function values for a single coordinate is always one, so the sum:
\begin{equation}\sum\limits_i \left( \sum\limits_{I} C_I(\vec{r}_i) \right) = n_e\end{equation}
is the total number of electrons $n_e$. Consequently:
\begin{equation} \sum\limits_{IJ} \frac{\partial\Phi}{\partial F_{IJ}} - n_e^2\Phi = \vec{0} \end{equation}
\begin{equation} \sum\limits_{J} \left( \frac{\partial\Phi}{\partial F_{IJ}} + \frac{\partial \Phi}{\partial F_{JI}}\right) - 2n_e\frac{\partial \Phi}{\partial G_{I}} = \vec{0} \end{equation}
As the zeroth-order wavefunction $\Phi$ is included in the linear method space, these are linear dependencies in the form of equation \ref{eq-lindep}. After integrating these constant linear coefficients, these supply us with the equivalence paths:
\begin{equation} \vec{p}_1(\alpha) = (\mat{F} + \alpha\doubleunderline{\mat{1}} , \vec{G}, \{g_I \})\end{equation}
\begin{equation} \vec{p}_{2,J}(\alpha) =   (\mat{F}  + \alpha(\underline{\vec{1}}^T\vec{e}_J + \vec{e}_J^T \underline{\vec{1}}), \vec{G} - \alpha (2n_e \vec{e}_J), \{g_I\}) \end{equation}
with:
\begin{equation}\doubleunderline{\mat{1}}_{ij} = 1, \quad \underline{\vec{1}}_{i} = 1, \quad (\vec{e}_j)_i = \delta{ij}
\end{equation}
We restrict variation along each of these paths by freezing the following parameter values:
\begin{equation} \vec{G} = \vec{0}, \quad F_{n_C, n_C} = 0, \quad n_C = \text{rank}(\mat{F}) \end{equation}
and omitting their parameter derivatives from the linear method tangent space.

Multiplying each of the anchor gaussians in $\{g_I\}$ by a single gaussian function $g$ produces the same normalized counting function basis $\{C_I\}$, as this operation is equivalent to multiplying the numerator and denominator of equation \ref{eq-normgauss} by $g$. 
Since multiplying two gaussian functions together produces another gaussian function, this directly gives the equivalence path:
\begin{equation} \vec{p}_3( g ) = (\mat{F}, \vec{G}, \{g_I / g \} )  \end{equation}
The coefficients $\vec{p}'_3(g)$ that explicitly appear in the associated linear dependencies are more complex, as they depend on the value of gaussian parameters internal to $g$. 
To remove these dependencies, we find an equivalent set of anchor gaussians with ten fewer parameters by dividing each by a single anchor gaussian $g_r$:
\begin{equation} \{\tilde{g}_I\} = \{1, g_2/g_r, \dots, g_{n}/g_r\} \end{equation}
and omit the parameters of $g_r$ from the linear method space. 

\section{Spherical Voronoi Partitioning}
\label{app-spherical_voronoi}

Basis functions written in terms of spherical coordinates are a natural choice in chemical systems, as discussed in section \ref{sec-spherical_voronoi}.
Here we provide an algorithm that generates counting functions that simultaneously divide space into angular sectors and radial shells.
First, we define an partition on the surface of a sphere with a set of unit vectors $\hat{\mu}_i$.
We will refer to anchor gaussians by the pair index $(j,i)$: 
\begin{equation} g_{(j,i)}(\vec{r}) =  \exp\left( -( \vec{r}^T \mat{A}_{(j,i)}\vec{r} - 2\vec{B}_{(j,i)}^T\vec{r} + K_{(j,i)} ) \right) \end{equation}
where $j$ indexes the radial shell (in order from smallest to largest radius) and $i$ indexes the angular position.
We then specify the maximum slope $S$ across switching surfaces set up by the angular partition, which happens to be consistent across radial shells, and set
\begin{equation} \alpha = S/d_{min}\end{equation}
where $d_{min}$ is the minimum Cartesian distance between the $\hat{\mu}_i$:
\begin{equation}d_{min} = \min\limits_{i\ne j} |\hat{\mu}_i - \hat{\mu}_j|\end{equation}
This gives the following parameters for the innermost shell of anchor gaussians:
\begin{equation} \mat{A}_{(0,i)} = \alpha \mat{I}, \quad \vec{B}_{(0,i)} = \alpha \hat{\mu}_i, \quad K_{(0,i)} = \alpha \end{equation}
The pair counting region boundaries between any pair of these anchor gaussians is a plane that intersects the center of the sphere, and partitions space into a set of angular sectors.
To further divide this partition along radial shells, we iteratively generate anchor gaussian parameters using those from the current outermost shell:
\begin{align}
\mat{A}_{(j,i)} &= \mat{A}_{(j-1,i)}\left(\frac{R_j - 2 S_{j}}{R_j}\right) \\
\vec{B}_{(j,i)} &= \vec{B}_{(j-1,i)} \\
K_{(j,i)} &= K_{(j-1,i)} - 2 S_j R_j
\end{align}
The pair counting function made from anchor gaussians on neighboring shells has a spherical switching boundary centered at the origin with radius $R_j$ and with maximum slope $S_j$, which may be verified by comparing to equation \ref{eq-quad_cf}:
\begin{equation}C_{(j,i), (j-1,i)}(\vec{r}) = \frac{1}{1 + \exp\left(-2R_jS_j \left( \frac{r^2}{R_j^2} - 1\right) \right)}\end{equation}

\section{Region Composition}
\label{app-region_composition}

In sections \ref{sec-classical_voronoi} and \ref{sec-spherical_voronoi} we describe how to partition space into atom-centered Voronoi cells and spherical Voronoi cells, and in section \ref{sec-region_composition}, we argue that subdividing regions within a single Jastrow basis is necessary to capture both interatomic and intraatomic population correlations simultaneously.
We do this by `composing' two existing, independent sets of counting functions into a single set while retaining the boundaries present in both partitions simultaneously as best we can. Conditions in equations \ref{eq-cond1} and \ref{eq-cond2} should be satisfied in this composition process and we provide a scheme that best meets these conditions here.

We will start by considering the case shown in figure \ref{fig-composition_labelled}, which starts with a rough partition of two counting functions ($C_{\alpha}$ and $C_{\beta}$) and a set of counting functions ($\{C_I\}$) that describe how $C_{\alpha}$ is to be subdivided.
Our desired result is a set of anchor gaussians $\{g_I^{(\alpha)}\}$ that, when used to replace the anchor gaussian $g_{\alpha}$ in the initial two-region partition, generates counting functions 
that reproduce the boundaries between the counting regions $\mathcal{R}_I$ within the region $\mathcal{R}_{\alpha}$:
\begin{align}
C_I^{(\alpha)}(\vec{r}) &= \frac{g_I^{(\alpha)}{\vec{r}}}{\sum\limits_J g_J^{(\alpha)}(\vec{r}) + C_{\beta}(\vec{r})} \notag \\
&= \frac{g_I(\vec{r})}{\sum\limits_J g_J(\vec{r})}, \quad \text{ whenever } \vec{r} \in \mathcal{R}_{\alpha}
\end{align}
and whose counting functions can be added together to reproduce the counting function $C_{\alpha}(\vec{r})$ everywhere:
\begin{equation} C_{\alpha}(\vec{r}) = \frac{g_{\alpha}(\vec{r})}{g_{\alpha}(\vec{r}) + g_{\beta}(\vec{r})} = \frac{\sum\limits_{J} g_J^{(\alpha)}(\vec{r}) }{ \sum\limits_{J} g_J^{(\alpha)}(\vec{r})  + g_{\beta}(\vec{r}) }\end{equation}
The first condition will be satisfied if we set:
\begin{equation}\label{eq-comp_prop_1} g_I^{(\alpha)} = g_I(\vec{r}) \cdot g_{r}(\vec{r})\end{equation}
for any gaussian function $g_r(\vec{r})$, since the boundaries between the counting regions of $C_I^{(\alpha)}$ are characterized by the pair counting functions:
\begin{equation} C_{I/J}^{(\alpha)}(\vec{r}) = \frac{g_I^{(\alpha)}(\vec{r})}{\sum\limits_J g_J^{(\alpha)}(\vec{r})} = \frac{g_I(\vec{r})}{\sum\limits_J g_J(\vec{r})} = C_{I/J}(\vec{r}) \end{equation}
which are the same as those of the original set $\{C_I\}$.

To satisfy the second condition, we substitute equation \ref{eq-comp_prop_1} into the expression for $C_{\alpha}(\vec{r})$: 
\begin{equation} C_{\alpha}(\vec{r}) = \frac{g_{\alpha}(\vec{r}) }{g_{\alpha}(\vec{r}) + g_{\beta}(\vec{r})} = \frac{\sum\limits_{J} g_J(\vec{r}) g_r(\vec{r}) }{ \left(\sum\limits_{J} g_J(\vec{r})\right) g_r(\vec{r}) + g_{\beta}(\vec{r}) }\end{equation}
After solving for $g_r(\vec{r})$, this becomes:
\begin{equation} \label{eq-comp_match} g_r(\vec{r}) = \frac{g_{\alpha}(\vec{r})}{\sum\limits_J g_J(\vec{r})}\end{equation}
The mismatch of the functional forms on the each side of this equation means that a solution for the gaussian parameters of $g_r$ that holds for all $\vec{r}$ will not be possible except in trivial or limiting cases.
We instead match a second-order Taylor expansion of the natural logarithm of each side at a single point $\vec{r}_0$:
\begin{align}
\ln(g_r(\vec{r})) &=  K_r - 2\vec{r}^T \vec{B}_r + \vec{r}^T \mat{A}_r \vec{r} \notag \\
&=  f(\vec{r}_0) + (\vec{r} - \vec{r}_0) \cdot \nabla f\big|_{\vec{r}_0} \notag \\
& \hspace{15mm} + \frac{1}{2}\left(\vec{r} - \vec{r}_0\right)^T \mat{H}\big|_{\vec{r}_0} \left(\vec{r}- \vec{r}_0\right)
\label{eq-log_gauss}
\end{align}
Where $f(\vec{r}), \nabla f\big|_{\vec{r}}, \mat{H}\big|_{\vec{r}}$ are the right-hand side of equation \ref{eq-comp_match}, its gradient, and its Hessian, evaluated at $\vec{r}$. Matching terms of the same order on each side gives us an explicit expression for each gaussian parameter in $g_r$:
\begin{equation} \label{eq-refsolve_const}K_r = \frac{1}{2}\vec{r}_0^T \mat{H}\big|_{\vec{r}_0} \vec{r}_0 -  \nabla f\big|_{\vec{r}_0} \cdot \vec{r}_0 + f(\vec{r}_0) \end{equation}
\begin{equation}\label{eq-refsolve_lin} \vec{B}_r = \frac{\mat{H}\big|_{\vec{r}_0}\vec{r}_0 - \nabla f\big|_{\vec{r}_0} }{2}  \end{equation}
\begin{equation} \label{eq-refsolve_quad}\mat{A}_r = \mat{H}\big|_{\vec{r}_0} \end{equation}
In this case, where we are subdividing an atom-centered Voronoi region into spherical shells, we choose $\vec{r}_0$ as a point midway between neighboring atomic centers, directly on the region boundary. In our simple two-region example, this is:
\begin{equation} \vec{r}_0 = \frac{\vec{\mu}_{\alpha} + \vec{\mu}_{\beta}}{2}\end{equation}
This selection is made to best preserve the counting region boundary along the bond axis.

We can generalize this to replace both atom-centered regions $\mathcal{R}_{\alpha}$ and $\mathcal{R}_{\beta}$ with multiple subpartitions $\{\mathcal{R}_{I,\alpha}\}$ and $\{\mathcal{R}_{I,\beta}\}$ simultaneously. 
Working through exactly the same logic with the first condition provides a similar prescription for $g_I^{(\alpha)}$ and $g_I^{(\beta)}$:
\begin{equation}\label{eq-comp_2ref} g_I^{(\alpha)}(\vec{r}) = g_{I,\alpha}(\vec{r}) g_r^{(\alpha)}(\vec{r}), \quad g_I^{(\beta)}(\vec{r}) = g_{I,\beta}(\vec{r}) g_r^{(\beta)}(\vec{r})\end{equation}
We substitute this into the second condition, with:
\begin{equation} S_{\alpha}(\vec{r}) = \sum\limits_I g_{I,\alpha}(\vec{r}),\quad S_{\beta}(\vec{r}) = \sum\limits_I g_{I,\beta}(\vec{r})\end{equation}
which gives:
\begin{equation}C_{\alpha}(\vec{r}) = \frac{g_{\alpha}(\vec{r})}{g_{\alpha}(\vec{r}) + g_{\beta}(\vec{r})} = \frac{S_{\alpha}(\vec{r})g_{r}^{(\alpha)}}{S_{\alpha}(\vec{r})g_{r}^{(\alpha)} + S_{\beta}(\vec{r})g_{r}^{(\beta)}} \end{equation}
As we can divide both numerator and denominator on the right-hand side by $g_r^{(\alpha)}$, the behavior of the composed set of anchor gaussians is uniquely determined by the quotient of gaussians,
\begin{equation} g_q(\vec{r}) = \frac{g_r^{(\beta)}}{g_r^{(\alpha)}(\vec{r})}\end{equation}
 which we solve for as before:
\begin{equation} \label{eq-double_comp}g_q(\vec{r})  = \frac{g_r^{(\beta)} }{g_r^{(\alpha)}} = \frac{S_{\alpha}(\vec{r}) g_{\beta}(\vec{r})}{S_{\beta}(\vec{r})g_{\alpha}(\vec{r})}\end{equation}
We conclude exactly as before by matching the second-order Taylor series of the logarithms of both sides of equation \ref{eq-double_comp} to solve for the gaussian parameters of $g_q$.
In cases where performing the composition serially may distort the counting region boundary, this latter scheme more faithfully reproduces the original boundaries of $\mathcal{R}_{\alpha}$ and $\mathcal{R}_{\beta}$ in the final composed basis. 


\bibliographystyle{aip}


\begin{thebibliography}{10}

\bibitem{helgaker}
T.~Helgaker, P.~J{\o}rgensen, and J.~Olsen,
\newblock {\em Molecular Electronic Structure Theory},
\newblock John Wiley \& Sons, LTD, Chichester, 2000.

\bibitem{szabo_ostlund}
A.~Szabo and N.~S. Ostlund,
\newblock {\em Modern Quantum Chemistry: Introduction to Advanced Electronic
  Structure Theory},
\newblock Dover Publications, Inc., Mineola, first edition, 1996.

\bibitem{strong_weak_correlation}
A.~D. Becke,
\newblock The Journal of Chemical Physics {\bf 138}, 074109 (2013).

\bibitem{kato_cusps}
T.~Kato,
\newblock Communications on Pure and Applied Mathematics {\bf 10}, 151.

\bibitem{needs_cusp}
N.~D. Drummond, M.~D. Towler, and R.~J. Needs,
\newblock Phys. Rev. B {\bf 70}, 235119 (2004).

\bibitem{qmc_review}
W.~Foulkes, L.~Mitas, R.~Needs, and G.~Rajagopal,
\newblock Reviews of Modern Physics {\bf 73}, 33 (2001).

\bibitem{parsegian_vdw}
V.~A. Parsegian,
\newblock {\em Van der Waals forces: a handbook for biologists, chemists,
  engineers, and physicists},
\newblock Cambridge University Press, 2005.

\bibitem{valeev_f12}
L.~Kong, F.~A. Bischoff, and E.~F. Valeev,
\newblock Chemical Reviews {\bf 112}, 75 (2012),
\newblock PMID: 22176553.

\bibitem{pulay_active_space}
S.~Keller, K.~Boguslawski, T.~Janowski, M.~Reiher, and P.~Pulay,
\newblock The Journal of Chemical Physics {\bf 142}, 244104 (2015).

\bibitem{chromium_dimer}
S.~Vancoillie, P.~{\r{A}}. Malmqvist, and V.~Veryazov,
\newblock Journal of Chemical Theory and Computation {\bf 12}, 1647 (2016),
\newblock PMID: 26930185.

\bibitem{oec_dmrg}
Y.~Kurashige, G.~K.-L. Chan, and Y.~Takeshi,
\newblock Nature Chemistry {\bf 5}, 662 (2013).

\bibitem{lee2018open}
J.~Lee, D.~W. Small, and M.~Head-Gordon,
\newblock arXiv preprint arXiv:1808.06743  (2018).

\bibitem{hartree_casscf}
D.~R. Hartree, F.~R. S., W.~Hartree, and B.~Swirles,
\newblock Philosophical Transactions of the Royal Society of London A:
  Mathematical, Physical and Engineering Sciences {\bf 238}, 229 (1939).

\bibitem{roos_casscf}
B.~O. Roos,
\newblock {\em The Complete Active Space Self-Consistent Field Method and its
  Applications in Electronic Structure Calculations}, pages 399--445,
\newblock Wiley-Blackwell, 2007.

\bibitem{rasscf}
J.~Olsen, B.~O. Roos, P.~J{\o}rgensen, and H.~J.~A. Jensen,
\newblock The Journal of Chemical Physics {\bf 89}, 2185 (1988).

\bibitem{white_dmrg2}
S.~R. White,
\newblock Phys. Rev. Lett. {\bf 69}, 2863 (1992).

\bibitem{white_dmrg}
S.~R. White,
\newblock Phys. Rev. B {\bf 48}, 10345 (1993).

\bibitem{chan_dmrg}
G.~K.-L. Chan, A.~Keselman, N.~Nakatani, Z.~Li, and S.~R. White,
\newblock The Journal of Chemical Physics {\bf 145}, 014102 (2016).

\bibitem{chan_dmrg2}
S.~Guo, Z.~Li, and G.~K.-L. Chan,
\newblock The Journal of Chemical Physics {\bf 148}, 221104 (2018).

\bibitem{kawakami_dmrg}
T.~Kawakami et~al.,
\newblock Chemical Physics Letters {\bf 705}, 85  (2018).

\bibitem{huron_sci}
B.~Huron, J.~P. Malrieu, and P.~Rancurel,
\newblock The Journal of Chemical Physics {\bf 58}, 5745 (1973).

\bibitem{evangelisti_cipsi}
S.~Evangelisti, J.-P. Daudey, and J.-P. Malrieu,
\newblock Chemical Physics {\bf 75}, 91  (1983).

\bibitem{sharma_hci}
J.~E.~T. Smith, B.~Mussard, A.~A. Holmes, and S.~Sharma,
\newblock Journal of Chemical Theory and Computation {\bf 13}, 5468 (2017).

\bibitem{sharma_hci2}
S.~Sharma, A.~A. Holmes, G.~Jeanmairet, A.~Alavi, and C.~J. Umrigar,
\newblock Journal of Chemical Theory and Computation {\bf 13}, 1595 (2017),
\newblock PMID: 28263594.

\bibitem{giner_scidmc}
E.~Giner, A.~Scemama, and M.~Caffarel,
\newblock {Canadian Journal of Chemistry} {\bf 91}, 879 (2013).

\bibitem{giner_scidmc2}
E.~Giner, A.~Scemama, and M.~Caffarel,
\newblock The Journal of Chemical Physics {\bf 142}, 044115 (2015).

\bibitem{alavi_fciqmc}
G.~H. Booth, A.~J.~W. Thom, and A.~Alavi,
\newblock The Journal of Chemical Physics {\bf 131}, 054106 (2009).

\bibitem{alavi_fciqmc_nature}
G.~H. Booth, A.~Gr{\"u}neis, and A.~Alavi,
\newblock Nature {\bf 493}, 365 (2013).

\bibitem{foulkes_fciqmc}
J.~S. Spencer, N.~S. Blunt, and W.~M. Foulkes,
\newblock The Journal of Chemical Physics {\bf 136}, 054110 (2012).

\bibitem{andersson_caspt2}
K.~Andersson, P.~A. Malmqvist, B.~O. Roos, A.~J. Sadlej, and K.~Wolinski,
\newblock Journal of Physical Chemistry {\bf 94}, 5483 (1990).

\bibitem{lester_qmc}
B.~M. Austin, D.~Y. Zubarev, and W.~A. Lester,
\newblock Chemical Reviews {\bf 112}, 263 (2012),
\newblock PMID: 22196085.

\bibitem{nightingale_jastrow}
C.-J. Huang, C.~J. Umrigar, and M.~P. Nightingale,
\newblock The Journal of Chemical Physics {\bf 107}, 3007 (1997).

\bibitem{mood_jastrow}
A.~L{\"{u}}chow, A.~Sturm, C.~Schulte, and K.~Haghighi~Mood,
\newblock The Journal of Chemical Physics {\bf 142}, 084111 (2015).

\bibitem{needs_jastrow}
P.~L{\'{o}}pez~R{\'{i}}os, P.~Seth, N.~D. Drummond, and R.~J. Needs,
\newblock Phys. Rev. E {\bf 86}, 036703 (2012).

\bibitem{sorella_jastrow4}
M.~Casula and S.~Sorella,
\newblock The Journal of Chemical Physics {\bf 119}, 6500 (2003).

\bibitem{sorella_jastrow5}
M.~Casula, C.~Attaccalite, and S.~Sorella,
\newblock The Journal of Chemical Physics {\bf 121}, 7110 (2004).

\bibitem{sorella_jastrow}
S.~Sorella, M.~Casula, and D.~Rocca,
\newblock The Journal of Chemical Physics {\bf 127}, 014105 (2007).

\bibitem{beaudet_jastrows}
T.~D. Beaudet, M.~Casula, J.~Kim, S.~Sorella, and R.~M. Martin,
\newblock The Journal of Chemical Physics {\bf 129}, 164711 (2008).

\bibitem{guidoni_jastrow}
F.~Sterpone, L.~Spanu, L.~Ferraro, S.~Sorella, and L.~Guidoni,
\newblock Journal of Chemical Theory and Computation {\bf 4}, 1428 (2008),
\newblock PMID: 26621429.

\bibitem{sorella_jastrow2}
M.~Marchi, S.~Azadi, M.~Casula, and S.~Sorella,
\newblock The Journal of Chemical Physics {\bf 131}, 154116 (2009).

\bibitem{guidoni_jastrow2}
M.~Barborini, S.~Sorella, and L.~Guidoni,
\newblock Journal of Chemical Theory and Computation {\bf 8}, 1260 (2012),
\newblock PMID: 24634617.

\bibitem{sorella_jastrow3}
A.~Zen, Y.~Luo, G.~Mazzola, L.~Guidoni, and S.~Sorella,
\newblock The Journal of Chemical Physics {\bf 142}, 144111 (2015).

\bibitem{neuscamman_cjagp}
E.~Neuscamman,
\newblock Journal of Chemical Theory and Computation {\bf 12}, 3149 (2016),
\newblock PMID: 27281678.

\bibitem{linderberg_agp}
J.~Linderberg,
\newblock Israel Journal of Chemistry {\bf 19}, 93.

\bibitem{gutzwiller_gf}
M.~C. Gutzwiller,
\newblock Phys. Rev. {\bf 137}, A1726 (1965).

\bibitem{neuscamman_jagp}
E.~Neuscamman,
\newblock Physical review letters {\bf 109}, 203001 (2012).

\bibitem{vdg_ncjf}
B.~V.~D. Goetz and E.~Neuscamman,
\newblock Journal of chemical theory and computation {\bf 13}, 2035 (2017).

\bibitem{sethian_levelset}
J.~A. Sethian,
\newblock {\em Level set methods and fast marching methods: evolving interfaces
  in computational geometry, fluid mechanics, computer vision, and materials
  science}, volume~3,
\newblock Cambridge university press, 1999.

\bibitem{mclachlan_discriminant_analysis}
G.~McLachlan,
\newblock {\em Discriminant analysis and statistical pattern recognition},
  volume 544,
\newblock John Wiley \& Sons, 2004.

\bibitem{mclachlan_mixture_models}
G.~J. McLachlan and K.~E. Basford,
\newblock {\em Mixture models: Inference and applications to clustering},
  volume~84,
\newblock Marcel Dekker, 1988.

\bibitem{aurenhammer2013voronoi}
F.~Aurenhammer, R.~Klein, and D.-T. Lee,
\newblock {\em Voronoi diagrams and Delaunay triangulations},
\newblock World Scientific Publishing Company, 2013.

\bibitem{gamess1}
M.~W. Schmidt et~al.,
\newblock Journal of Computational Chemistry {\bf 14}, 1347.

\bibitem{gamess2}
M.~S. Gordon and M.~W. Schmidt,
\newblock Chapter 41 - advances in electronic structure theory: Gamess a decade
  later,
\newblock in {\em Theory and Applications of Computational Chemistry}, edited
  by C.~E. Dykstra, G.~Frenking, K.~S. Kim, and G.~E. Scuseria, pages 1167 --
  1189, Elsevier, Amsterdam, 2005.

\bibitem{MOLPRO-WIREs}
H.-J. Werner, P.~J. Knowles, G.~Knizia, F.~R. Manby, and M.~Sch{\"u}tz,
\newblock WIREs Comput Mol Sci {\bf 2}, 242 (2012).

\bibitem{MOLPRO}
H.-J. Werner et~al.,
\newblock Molpro, version 2018.1, a package of ab initio programs, 2018,
\newblock see http://www.molpro.net.

\bibitem{molpro_integrals}
R.~Lindh, U.~Ryu, and B.~Liu,
\newblock The Journal of chemical physics {\bf 95}, 5889 (1991).

\bibitem{molpro_ccsd_mp2}
C.~Hampel, K.~A. Peterson, and H.-J. Werner,
\newblock Chemical physics letters {\bf 190}, 1 (1992).

\bibitem{molpro_uccsd}
C.~Knowles, Peter J .and~Hampel and H.-J. Werner,
\newblock The Journal of chemical physics {\bf 99}, 5219 (1993).

\bibitem{molpro_ccsdt}
M.~J. Deegan and P.~J. Knowles,
\newblock Chemical physics letters {\bf 227}, 321 (1994).

\bibitem{molpro_mcscf1}
H.-J. Werner and P.~J. Knowles,
\newblock The Journal of Chemical Physics {\bf 82}, 5053 (1985).

\bibitem{molpro_mcscf2}
P.~J. Knowles and H.-J. Werner,
\newblock Chemical Physics Letters {\bf 115}, 259 (1985).

\bibitem{molpro_mrci1}
H.-J. Werner and P.~J. Knowles,
\newblock The Journal of Chemical Physics {\bf 89}, 5803 (1988).

\bibitem{molpro_mrci2}
P.~J. Knowles and H.-J. Werner,
\newblock Chemical Physics Letters {\bf 145}, 514 (1988).

\bibitem{molpro_mrci3}
K.~Shamasundar, G.~Knizia, and H.-J. Werner,
\newblock The Journal of chemical physics {\bf 135}, 054101 (2011).

\bibitem{dmc_core_scaling}
B.~L. Hammond, P.~J. Reynolds, and W.~A. Lester,
\newblock The Journal of Chemical Physics {\bf 87}, 1130 (1987).

\bibitem{carbon_oxygen_ecp}
A.~Bergner, M.~Dolg, W.~K{\"u}chle, H.~Stoll, and H.~Preu{\ss},
\newblock Molecular Physics {\bf 80}, 1431 (1993).

\bibitem{calcium_ecp}
M.~Kaupp, P.~v.~R. Schleyer, H.~Stoll, and H.~Preuss,
\newblock The Journal of chemical physics {\bf 94}, 1360 (1991).

\bibitem{qmcpack}
J.~Kim et~al.,
\newblock Journal of Physics: Condensed Matter {\bf 30}, 195901 (2018).

\bibitem{linearmethod1}
J.~Toulouse and C.~J. Umrigar,
\newblock The Journal of Chemical Physics {\bf 126}, 084102 (2007).

\bibitem{linearmethod2}
J.~Toulouse and C.~J. Umrigar,
\newblock The Journal of Chemical Physics {\bf 128}, 174101 (2008).

\bibitem{dunning_basis}
T.~H. Dunning,
\newblock The Journal of Chemical Physics {\bf 90}, 1007 (1989).

\bibitem{varandas_basisextrap}
A.~J.~C. Varandas,
\newblock The Journal of Chemical Physics {\bf 126}, 244105 (2007).

\bibitem{gaussian_opt}
R.~E. Kari, P.~G. Mezey, and I.~G. Csizmadia,
\newblock The Journal of Chemical Physics {\bf 63}, 581 (1975).

\bibitem{chandrasekharan_signproblem}
S.~Chandrasekharan and U.-J. Wiese,
\newblock Physical Review Letters {\bf 83}, 3116 (1999).

\bibitem{nlpp_locality_ceperley}
L.~Mit{\'{a}}{\u{s}}, E.~L. Shirley, and D.~M. Ceperley,
\newblock The Journal of Chemical Physics {\bf 95}, 3467 (1991).

\bibitem{casula_tmoves}
M.~Casula,
\newblock Phys. Rev. B {\bf 74}, 161102 (2006).

\bibitem{nlpp_tmoves_filippi}
M.~Casula, S.~Moroni, S.~Sorella, and C.~Filippi,
\newblock The Journal of Chemical Physics {\bf 132}, 154113 (2010).

\bibitem{nlppe_benchmark}
R.~Nazarov, L.~Shulenburger, M.~Morales, and R.~Q. Hood,
\newblock Physical Review B {\bf 93}, 094111 (2016).

\bibitem{sergio_multislater}
S.~D. {Pineda Flores} and E.~{Neuscamman},
\newblock ArXiv e-prints  (2018).

\bibitem{claudia_multislater}
R.~Assaraf, S.~Moroni, and C.~Filippi,
\newblock Journal of Chemical Theory and Computation {\bf 13}, 5273 (2017),
\newblock PMID: 28873307.

\bibitem{bigdmc_1}
E.~R. Batista et~al.,
\newblock Phys. Rev. B {\bf 74}, 121102 (2006).

\bibitem{bigdmc_2}
J.~Yu, L.~K. Wagner, and E.~Ertekin,
\newblock The Journal of Chemical Physics {\bf 143}, 224707 (2015).

\bibitem{bigdmc_3}
H.~Zheng and L.~K. Wagner,
\newblock Phys. Rev. Lett. {\bf 114}, 176401 (2015).

\bibitem{zhao2017blocked}
L.~Zhao and E.~Neuscamman,
\newblock Journal of chemical theory and computation {\bf 13}, 2604 (2017).

\bibitem{shea_excited}
J.~A.~R. Shea and E.~Neuscamman,
\newblock Journal of Chemical Theory and Computation {\bf 13}, 6078 (2017),
\newblock PMID: 29140699.

\end{thebibliography}

\end{document}